\documentclass[10pt,conference]{IEEEtran}
\usepackage{amsmath,amssymb,amsfonts}
\usepackage{algorithmic}
\usepackage{graphicx}
\usepackage{xcolor}
\usepackage[hyphens]{url}
\usepackage{fancyhdr}

\usepackage{hyperref}
\usepackage[most]{tcolorbox}
\usepackage{arydshln} 
\usepackage{lineno}
\usepackage{soul}
\usepackage{bbding}
\usepackage{wasysym}
\usepackage{textcomp}
\usepackage{tabularx}
\usepackage{hyperref}
\usepackage{threeparttable}
\usepackage{multirow}
\usepackage[pass]{geometry}
\usepackage{colortbl}
\usepackage{pifont}
\usepackage{bm}
\usepackage{utfsym}
\usepackage{arydshln}
\usepackage{lineno}
\usepackage{booktabs}
 
\usepackage{marvosym}

\hypersetup{
  colorlinks=true,
  linkcolor=blue,
  urlcolor=blue,
  citecolor=blue
}

\tcbset{observationbox/.style={
  colframe=black, colback=blue!2,
  boxrule=0.7pt, arc=1pt,
  left=4pt, right=4pt, top=0pt, bottom=0pt,
  fonttitle=\bfseries
}}

\tcbset{guidancebox/.style={
  colframe=black, colback=gray!2,
  boxrule=0.7pt, arc=1pt,
  left=4pt, right=4pt, top=0pt, bottom=0pt,
  fonttitle=\bfseries
}}

\tcbset{Challenge/.style={
  colframe=black, colback=red!2,
  boxrule=0.7pt, arc=1pt,
  left=4pt, right=4pt, top=-0.2pt, bottom=-0.2pt,
  fonttitle=\bfseries
}}

\definecolor{mygray}{gray}{.94}
\definecolor{Common}{rgb}{0.9765, 0.9373, 0.7490}
\definecolor{RA}{rgb}{0.8471, 0.8980, 0.9765}
\definecolor{RB}{rgb}{0.8569,0.8659, 0.8667}
\definecolor{RC}{rgb}{0.9333,0.6471,0.6510}
\definecolor{RD}{rgb}{0.0353,1.0000,0.6000}
\definecolor{RE}{rgb}{0.9882,0.65,0.45}
\definecolor{RF}{rgb}{0.9882,0.8471,0.8039}

\newcommand{\hpcayear}{2026}

\title{PADE: A Predictor-Free Sparse Attention Accelerator via Unified Execution and Stage Fusion}

\def\hpcacameraready{} 

\newcommand\hpcaauthors{Huizheng Wang$^\dagger$, Hongbin Wang$^\dagger$, Zichuan Wang$^\dagger$, Zhiheng Yue$^\dagger$, Yang Wang$^\dagger$, Chao Li$^\ddagger$, Yang Hu$^\dagger$\textsuperscript{\Letter}, Shouyi Yin$^\dagger$$^*$}
\newcommand\hpcaaffiliation{$^\dagger$School of Integrated Circuits, BNRist, Tsinghua University, Beijing, China, 100084 \\
$^\ddagger$School of Computer Science and Engineering, Shanghai Jiao Tong University, Shanghai, China, 200240 \\
$^*$Shanghai Artificial Intelligence Laboratory, Shanghai, China, 200433}
\newcommand\hpcaemail{\textsuperscript{\Letter}Corresponding author, hu\_yang@tsinghua.edu.cn}

\author{
  \ifdefined\hpcacameraready
    \IEEEauthorblockN{\hpcaauthors{}}
      \IEEEauthorblockA{
        \hpcaaffiliation{} \\
        \hpcaemail{}
      }
  \else
    \IEEEauthorblockN{\normalsize{ 
    \colorbox{Common}{Common concerns},\colorbox{RA}{Reviewer A}, \colorbox{RB}{Reviewer B}, \colorbox{RC}{Reviewer C}, \colorbox{RD}{Reviewer D}, \colorbox{RE}{Reviewer E}, \colorbox{RF}{Reviewer F}
    } \\
    }
  \fi 
}

\fancypagestyle{camerareadyfirstpage}{%
  \fancyhead{}
  
  \fancyhead[C]{
    \ifdefined\aeopen
    \parbox[][12mm][t]{13.5cm}{\hpcayear{} IEEE International Symposium on High-Performance Computer Architecture (HPCA)}    
    \else
      \ifdefined\aereviewed
      \parbox[][12mm][t]{13.5cm}{\hpcayear{} IEEE International Symposium on High-Performance Computer Architecture (HPCA)}
      \else
      \ifdefined\aereproduced
      \parbox[][12mm][t]{13.5cm}{\hpcayear{} IEEE International Symposium on High-Performance Computer Architecture (HPCA)}
      \else
      \parbox[][0mm][t]{13.5cm}{\hpcayear{} IEEE International Symposium on High-Performance Computer Architecture (HPCA)}
    \fi 
    \fi 
    \fi 
    \ifdefined\aeopen 
      \includegraphics[width=12mm,height=12mm]{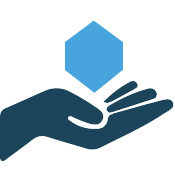}
    \fi 
    \ifdefined\aereviewed
      \includegraphics[width=12mm,height=12mm]{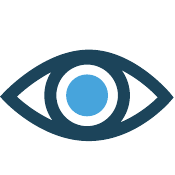}
    \fi 
    \ifdefined\aereproduced
      \includegraphics[width=12mm,height=12mm]{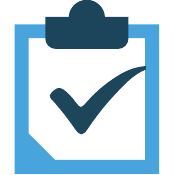}
    \fi
  }
  \fancyfoot[C]{}
}
\begin{document}
\maketitle

\ifdefined\hpcacameraready 
  \thispagestyle{camerareadyfirstpage}
  \pagestyle{empty}
\else
  \thispagestyle{plain}
  \pagestyle{plain}
\fi

\newcommand{\hpcaheight}{0mm}
\ifdefined\eaopen
\renewcommand{\hpcaheight}{12mm}
\fi



\begin{abstract}
Attention-based models have revolutionized AI, but the quadratic cost of self-attention incurs severe computational and memory overhead. Sparse attention methods alleviate this by skipping low-relevance token pairs. However, current approaches lack practicality due to the heavy expense of added sparsity predictor, which severely drops their hardware efficiency.

This paper advances the state-of-the-art (SOTA) by proposing a bit-serial enable stage-fusion (BSF) mechanism, which eliminates the need for a separate predictor. However, it faces key challenges: 1) Inaccurate bit-sliced sparsity speculation leads to incorrect pruning; 2) Hardware under-utilization due to fine-grained and imbalanced bit-level workloads. 3) Tiling difficulty caused by the row-wise dependency in sparsity pruning criteria. 

We propose PADE, a predictor-free algorithm-hardware co-design for dynamic sparse attention acceleration. PADE features three key innovations: 1) Bit-wise uncertainty interval-enabled guard filtering (BUI-GF) strategy to accurately identify trivial tokens during each bit round; 2) Bidirectional sparsity-based out-of-order execution (BS-OOE) to improve hardware utilization; 3) Interleaving-based sparsity-tiled attention (ISTA) to reduce both I/O and computational complexity. These techniques, combined with custom accelerator designs, enable practical sparsity acceleration without relying on an added sparsity predictor. Extensive experiments on 22 benchmarks show that PADE achieves $7.43\times$ speed up and $31.1\times$ higher energy efficiency than Nvidia H100 GPU. Compared to SOTA accelerators, PADE achieves $5.1\times$, $4.3\times$ and $3.4\times$ energy saving than Sanger, DOTA and SOFA.
\end{abstract}

\section{Introduction}
Transformer models have achieved significant success in various fields, spanning from content generation \cite{song2018situ,brown2020language,hendrycks2020measuring,lin2023videodirectorgpt} to computer vision \cite{dosovitskiy2020image,you2023vitcod,bai2024seed}. However, the self-attention mechanism used in Transformers suffers from quadratic time and memory complexity, limiting their scalability to long sequences.



To address this, \textit{sparse attention} techniques have emerged as a promising solution, where attention is computed over a subset of query-key (Q-K) pairs instead of a dense attention matrix. Existing sparse attention works can be divided into two routes: \emph{static sparsity (SS)} and \emph{dynamic sparsity (DS)}. SS \cite{parmar2018image,child2019generating,qiu2019blockwise,zaheer2020big,beltagy2020longformer,jiang2024minference,dong2022cswin,katharopoulos2020transformers} relies on predefined sparse patterns that remain fixed during inference, lacking flexibility and often resulting in significant accuracy degradation \cite{li2024large,fuad2023survey,kachris2025survey}. In contrast, DS \cite{correia2019adaptively,kitaev2020reformer,you2023vitcod,wang2023cta,song2024tsacc,cho2024sparc,park2024token,fan2022adaptable,dass2023vitality,li2020ftrans,wang2024sofa,qin2023fact,ham20203,ham2021elsa,lu2021sanger,qu2022dota,yang2022dtatrans,zhou2022energon,wang2025bitstopper,wang2021spatten,liu2022dynamic,wang2025lapa,li2022accelerating,wang2025mcbp} adapts the sparsity pattern at runtime, offering improved accuracy and flexibility, making it more suitable for a broader range of tasks and input types.

\begin{figure}[t]
\centering
\includegraphics[width=\linewidth]{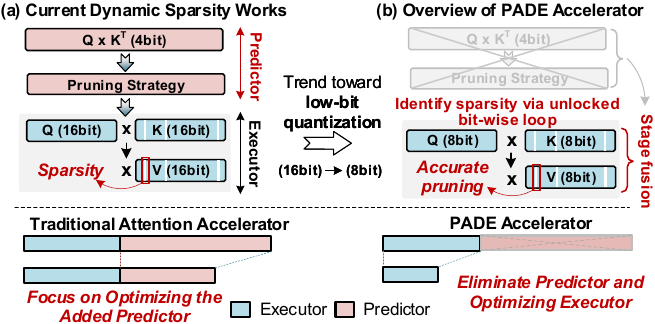}\vspace{-3mm}
\caption{Comparison of (a) current DS works and (b) PADE.}
\label{fig:DS}\vspace{-5mm}
\end{figure}

\textbf{However, such flexibility of DS comes at the expense of an additional sparsity predictor.} Fig. \ref{fig:DS} (a) outlines the typical workflow of existing DS attention accelerators \cite{ham20203,ham2021elsa,lu2021sanger,qu2022dota,zhou2022energon,yang2022dtatrans,cho2024sparc,liu2022dynamic,li2022accelerating,wang2021spatten,qin2023fact,wang2024sofa}, which consists of three stages. First, attention scores ($\mathbf{Q}$$\times$$\mathbf{K}^T$) are estimated via low-overhead techniques, such as 4-bit MSB multiplication \cite{lu2021sanger,zhou2022energon}, log-domain shifting \cite{qin2023fact,wang2024sofa}, low-rank approximation \cite{ham2021elsa,qu2022dota}, and clustering \cite{cho2024sparc,lee2025clat}. Next, a pruning strategy, like threshold comparison \cite{lu2021sanger,li2022accelerating,zhou2022energon} or top-$k$ sorting \cite{wang2021spatten,qin2023fact,wang2024sofa}, generates a sparsity mask for important QK pairs (iQKs). This process relies on an additional sparsity predictor. Finally, only iQKs are processed by the attention executor with higher bit-width precision (typically 16-bit), while ineffective QK-pairs (iEQKs) are directly pruned. 

\begin{figure}[t]
\centering
\includegraphics[width=\linewidth]{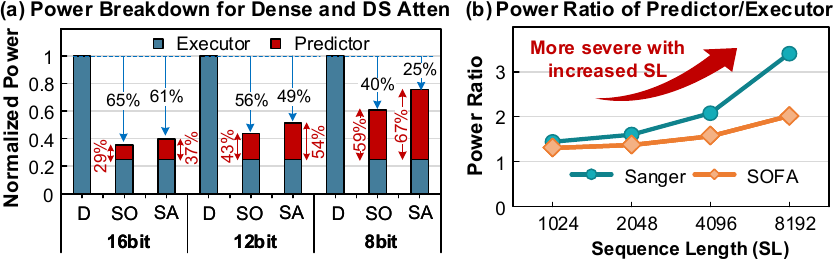}\vspace{-3mm}
\caption{(a) Power breakdown of dense and DS attention (SA: Sanger, SO: SOFA) with TSMC 28nm across executor bit-widths of Llama7B. (b) Power ratio of predictor and executor versus SL with under 8-bit quantized executor.}
\label{fig:Predictor_Power}\vspace{-4mm}
\end{figure}

\textbf{Unfortunately, such an added predictor occupies substantial overhead, increasingly offsetting the sparsity benefit}. Fig. \ref{fig:Predictor_Power} (a) shows the power breakdown of dense attention and two representative DS accelerators: Sanger \cite{lu2021sanger}, SOFA \cite{wang2024sofa}, with varying executor bit-widths. Power consumption is categorized into executor and predictor components. Sanger uses 4-bit MSB multiplication with threshold comparison, while SOFA employs log domain shifting with top-$k$ sorting. As depicted in Fig. \ref{fig:Predictor_Power}(a), there are two key observations: (\textbf{1}) At larger executor bit-widths (e.g., 16bit), the DS reduces overall power by about $63\%$, with the predictor costing only $33\%$. This explains why previous DS works adopt additional sparsity predictors. (\textbf{2}) However, as executor bit-width decreases, predictor overhead becomes dominant. At 8-bit, overall savings drop to merely $32\%$, with the sparsity predictor occupying over $63\%$ of total cost. This is due to the predictor must access and process full-sized K tensors, a cost unaffected by sparsity.

Further, Fig. \ref{fig:Predictor_Power}(b) reveals the predictor-to-executor power ratio under varying sequence lengths (SL). As can be seen, as the SL increases, this ratio grows noticeably for both designs, indicating the growing relative overhead of the predictor. This is because the increased sparsity in longer sequences exacerbates the predictor’s relative overhead.

\textbf{Takeaway}: With the rapid advancement of Transformer quantization techniques, like GPTQ \cite{frantar2022gptq}, LLM.int8() \cite{dettmers2022gpt3}, SmoothQuant \cite{xiao2023smoothquant} and Atom \cite{zhao2024atom}, there is a growing trend of adopting low-bit quantization in attention mechanisms. In these cases, the predictor's overhead increasingly offsets the benefits of sparsity. \textbf{This highlights a need to reduce or even eliminate the prediction overhead.}

\textbf{Insights: The root cause of the excessive prediction cost stems from the decoupling between existing sparsity predictors and executors}, which hinders the computational and memory access efforts paid in the predictor from being reused by the executor. To address this, we draw inspiration from bit-serial computing \cite{judd2016stripes,albericio2017bit,lee2018unpu,chen2024bbs,kam2025panacea,im2023sibia,guo2025transitive}, which separates an INT operation into multi-round bit-level steps. This motivates a unified design that integrates prediction and execution into a single computation stage, thereby eliminating the separate predictor and improving overall efficiency.

To realize this idea, we propose a bit-serial-enable stage fusion (BSF) strategy that eliminates the additional prediction stage via the following key steps: 1) Start with the first bit plane (i.e., MSB) of Keys for bit-serial speculating of $\mathbf{Q}\times \mathbf{K}^T$. 2) Once a token (Key) is identified as unlikely to be an important QK pair (i.e., iQK), its processing and associated memory access with subsequent bit planes are immediately terminated. In this way, the accelerator only needs to perform the remaining computation for the iQK, and obtains the final result by adding it to the previously generated partial result.  

Despite its potential, realizing a BSF-style DS accelerator presents several challenges: \textbf{(1)} Lack of an effective bit-wise decision mechanism for early iEQKs identification. \textbf{(2)} Hardware under-utilization due to fine-grained and imbalanced bit-level workloads. \textbf{(3)} Tiling difficulty arising from row-wise dependency in sparsity pruning criteria.




To this end, we propose PADE, a software-hardware co-design, whose high-level overview is depicted in Fig.\ref{fig:DS} (b). It features three key techniques that correlate to three challenges:

1) We propose Bit-level Uncertainty Interval-enabled Guarded Filtering (BUI-GF) to accurately identify iEQKs at each bit round. By exploiting properties of two’s complement representation, we define a Bit-wise Uncertainty Interval (BUI) to bound the potential variation of inner products. This safety margin enables precise and reliable early pruning decisions.

\begin{figure}[t]
\centering
\includegraphics[width=\linewidth]{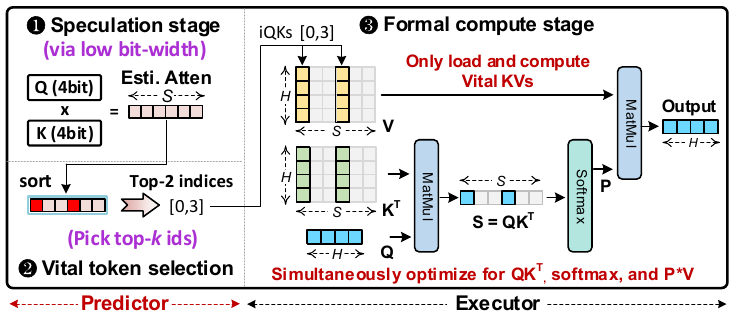}\vspace{-3mm}
\caption{Illustration of the DS attention mechanism.}
\label{fig:attention_mechanism}\vspace{-4mm}
\end{figure}

\begin{figure*}[t]
\centering
\includegraphics[width=\linewidth]{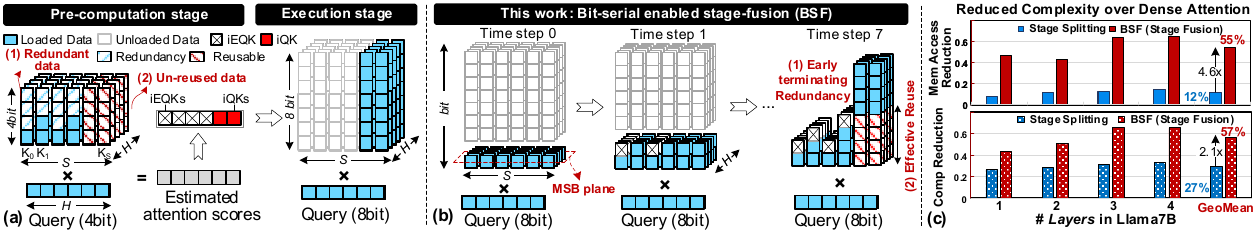}\vspace{-3.5mm}
\caption{(a) Traditional DS works, featuring stage splitting. (b) Our work features stage-fusion. (c) Reduced complexity for stage splitting and stage fusion.}
\label{fig:stage_fusion}\vspace{-2mm}
\end{figure*}

2) We propose a Bidirectional Sparsity-based Out-of-order Execution (BS-OOE) to improve bit-grained hardware utilization. It first introduces a bit-level bidirectional sparsity (BS) scheme to promote load balancing across PEs. Further, it utilizes out-of-order execution to hide DRAM access latency by breaking the constraint of conventional bit-serial computation. 

3) We propose an Interleaving-based Sparsity-Tiled Attention (ISTA) mechanism to improve IO efficiency. By exploiting the monotonicity of Softmax and the early termination property of bit-serial computation, we skillfully decompose pruning decisions to the tiling level. Additionally, an interleaved update strategy further minimizes redundant operations across tiles.

To support the above optimization mechanisms effectively, we
design a dedicated accelerator named PADE: \textbf{1)} For BUI-GF, it employs the dedicated scoreboard-based, result-reusable PE lane to eliminate redundant memory access across bit execution rounds, thus significantly reducing the energy overhead of repeatedly loading bit planes. \textbf{2)} For BS-OOE, PADE integrates the grouped, lightweight sparsity ANDer trees to mitigate the overhead of large multiplexers. \textbf{3)} For ISTA, a reuse-aware reorder scheduler is dedicated to improving tiling execution efficiency, by minimizing redundant memory access. The PADE accelerator achieves an average energy efficiency of 11740 GOPS/W, which is 31.1×, 5.1×, 4.3× and 3.4× higher than H100 GPU, SOTA accelerator Sanger, DOTA and SOFA.


\section{Background}\label{sec:background}
\subsection{Transformer and DS Attention}
\textbf{Transformer models}. Initially, the Transformer maps a length-
$S$ sequence into Q, K, and V spaces. Next, Q and K are multiplied to generate an attention score $\mathbf{S}$ with $\mathbb{R}^{S\times S}$, which captures token-to-token correlations. The $\mathbf{S}$ is then passed through a \textit{softmax} and multiplied with V activation, resulting in a matrix $\mathbf{O}\in \mathbb{R}^{S\times H}$, where $H$ denotes hidden dimension. Finally, a feed-forward network (FFN) generates the outputs.



\textbf{Dynamic Sparsity (DS) attention}. Typically, dense attention involves load and compute all Ks and Vs. In contrast, the DS attention fetches and processes only the important iQK pairs. As depicted in Fig. \ref{fig:attention_mechanism}, only the 0th and 3rd KVs are loaded and computed with the Query, thus effectively reducing the computation and memory access. However, this reduced complexity comes at the cost of an additional sparsity predictor, which involves low-bit QK speculation and vital token selection processes. This introduces non-negligible overhead. 

\section{Motivation}\label{sec:motivation}
\subsection{Re-examinating DS Works and Opportunity}
As shown in Fig. \ref{fig:stage_fusion}(a), traditional DS works adopt a \emph{stage-splitting} paradigm that decouples prediction from execution. Taking Sanger \cite{lu2021sanger} as an example, the predictor takes the full 4-bit Key tensor to identify iQKs, while the executor separately fetches the corresponding KVs for precise computation. \textbf{However, this paradigm introduces two inefficiency sources}: (\textbf{1}) For Keys related to iEQKs (e.g., K$_0$), a single bit may suffice to identify their insignificance. However, the stage-splitting approach blindly loads 4 bits, resulting in redundancy. (\textbf{2}) For Keys related to iQKs (e.g., K$_s$), the executor reloads their high-bit-width versions for subsequent more precise computation. However, it fails to reuse the data already processed during prediction, leading to inefficiency.

\textbf{Quantifying predictor overhead}. As shown in Fig. \ref{fig:Predictor_Power}, after sparsification, the added predictor incurs over 63\% power overhead, limiting the gain of DS attention to just $1.5\times$ over dense attention. To this end, we derive the design guidance:


\vspace{3mm}

\hspace{-3.5mm}
\fbox{%
  \begin{minipage}{0.96\linewidth}
  \vspace{0.8pt}
\textbf{(Design Guidance)} An ideal DS accelerator should eliminate the extra sparsity predictor while preserving sparsity.
  \end{minipage}
}

\vspace{3mm}

Motivated by this insight, the BSF strategy aims to unify prediction and execution within a single computation stage, thereby minimizing both computation and memory access. As illustrated in Fig. \ref{fig:stage_fusion}(b), BSF progressively extracts lower-order bit-planes to assess token importance during QK computation. This enables not only early termination for iEQKs but also computation and memory access reuse for iQKs.

Unlike stage-splitting DS designs that rely on separate predictors, BSF eliminates prediction overhead and enables fine-grained early termination to reduce unnecessary computation and memory access. Fig. \ref{fig:stage_fusion}(c) profiles the memory access and computation reduction achieved by different strategies in the attention modules across four layers from LLaMA-2\,7B \cite{touvron2023llama2}. On average, BSF can achieve $4.6\times$ higher memory access and $2.1\times$ more computation reduction, compared to traditional stage-splitting DS approaches, highlighting its significant energy efficiency advantage over current DS methods.

\begin{figure*}[t]
\centering
\includegraphics[width=\linewidth]{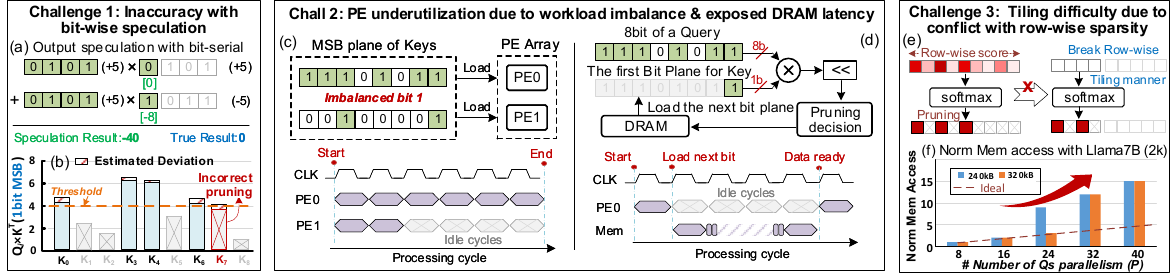}\vspace{-3mm}
\caption{Challenges for bit-serial enable stage fusion. (a)-(b) Inaccuracy (c)-(d) Hardware under-utilization. (e)-(f) Tiling difficulty. }
\label{fig:challlenge}\vspace{-4mm}
\end{figure*}

\subsection{Challenges for Bit-serial Enable Stage-fusion}\label{subsec:Challenges}
Despite its theoretical benefits, a naive implementation of BSF will encounter three challenges, as depicted in Fig. \ref{fig:challlenge}. 


\vspace{3mm}

\hspace{-3.5mm}
\fbox{%
  \begin{minipage}{0.96\linewidth}
  \vspace{0.8pt}
\textbf{(Challenge 1)} Incorrect pruning decisions caused by the inherent inaccuracy in bit-wise speculation.
  \end{minipage}
}

\vspace{3mm}

As shown in the Fig. \ref{fig:challlenge} (a), the 1-bit MSB representation of (+5) and 1-bit representation of (-5) are used to predict the result of (+5) $\times$ (+5) + (+5) $\times$ (-5). The true result should be 0. However, under the 1-bit representation, the MSB bit plane of (1011)$_2$ (-5) and (0101)$_2$ (+5) are regarded as (1000)$_2$ (-8) and (0000)$_2$ (+0), respectively. This leads to an estimated result of -40, which significantly deviates from the correct result. This severe error can lead to incorrect token pruning during early termination. For example, in K$_7$ of Fig. \ref{fig:challlenge} (b).


\textbf{Key idea}. Inspired the conservative margin concept \cite{li2022accelerating,park2024token}, we propose bit uncertainty interval-enabled guarded filtering (BUI-GF), a lightweight mechanism for simple yet accurate max-based pruning. BUI-GF characterizes the maximum possible fluctuation of dot products across bit planes using simple yet efficient bit flipping, enabling conservative yet hardware-efficient pruning decisions with minimal overhead. To support this, we design a scoreboard-based result-reuse PE to reduce BUI hardware overhead and redundant bit-plane memory accesses.


\vspace{3mm}

\hspace{-3.5mm}
\fbox{%
  \begin{minipage}{0.96\linewidth}
  \vspace{0.8pt}
\textbf{(Challenge 2)} Compute resource underutilization due to workload imbalance and exposed memory access latency.
  \end{minipage}
}

\vspace{3mm}

As depicted in Fig. \ref{fig:challlenge} (c), each bit-plane of different Keys is assigned to a separate PE for parallel execution. However, PEs corresponding to Keys whose bit-planes contain more `1’ bits will require more processing cycles (e.g., PE\,0), while those with fewer `1’ bits will finish earlier (e.g., PE\,1). This imbalance leads to computation stalls and under-utilization.

As shown in Fig. \ref{fig:challlenge} (d), to enable bit-grained early termination, it is critical to avoid the bulk loading of all bit-planes for each Key. Instead, each Key should be evaluated bit-plane by bit-plane to determine whether it qualifies as an iEQK. If true, the subsequent bit-planes for that Key are skipped; otherwise, the next bit-plane is required. However, due to DRAM's dynamic precharge mechanism, loading each bit-plane typically incurs several dozen cycles \cite{chang2016understanding,JEDEC_JESD235D}. Naively stalling computation during data loading results in underutilized computational resources.


\textbf{Key idea}. Inspired by bidirectional sparsity \cite{chen2024bbs} and XNOR-BNN-based formulations \cite{geng2019o3bnn}, we propose a two-pronged approach to mitigate resource underutilization. First, we introduce bidirectional, runtime-adaptive sparsity orchestration for the K matrix, which dynamically interprets bit ‘1’ as sparsity in coordination with queries, ensuring load imbalance remains below $50\%$. Building upon this, we further introduce bit-wise out-of-order (OOE) execution, allowing the PE to process other bit-planes while avoiding memory access stalls. To support this, we propose a temporal-reuse-based sparsity scheduler to alleviate runtime scheduling overhead, a lightweight ANDer tree to BS-induced multiplexing overhead, and a scoreboard-based PE that facilitates partial-sum buffering and reuse.

\vspace{3mm}

\hspace{-3.5mm}
\fbox{%
  \begin{minipage}{0.96\linewidth}
  \vspace{0.8pt}
\textbf{(Challenge 3)} IO inefficiency resulting from the conflict between tiling with the row-dependent pruning strategy.
  \end{minipage}
}

\vspace{3mm}


As shown in Fig. \ref{fig:challlenge} (e), existing pruning strategies rely on row-wise attention score distributions to assess token importance, introducing strong row dependencies. This dependency prevents effective tiling, which is essential for IO efficiency. Without tiling, when the number of parallel queries increases, memory access overhead grows sharply. As illustrated in Fig. \ref{fig:challlenge} (f), increasing the number of parallel queries ($P$) from 8 to 32, leads to over 12× more memory accesses. A coarse solution is to enlarge on-chip SRAM, but this incurs significant area inefficiency. For example, with ($P$=512, $S$=2048), 5MB of SRAM is required, resulting in a 5.47 mm$^2$ footprint under TSMC 28nm technology, which is $7.4\times$ and $8.9\times$ larger than the total area of SpAtten \cite{wang2021spatten} and ELSA \cite{ham2021elsa}, respectively.

\begin{table}[t]
\vspace{-2mm}
\renewcommand{\arraystretch}{0.97}
\caption{Summary for SOTA Attention Accelerators.}\vspace{-2.5mm}
\centering
\begin{threeparttable}
\footnotesize
\begin{tabular}{l||m{1.3cm}<{\centering}|m{1.0cm}<{\centering}|m{0.78cm}<{\centering}|m{0.8cm}<{\centering}|m{0.8cm}<{\centering}}
\specialrule{0.12em}{0.5pt}{0.4pt}

 \multirow{2}{*}{\!\!\!\textbf{Accelerator}} &   \multicolumn{2}{c|}{\textbf{Optimization}} &  {\!\!\!\!\textbf{Predictor}}\!\!\! & \textbf{Tiling} & {\!\!\textbf{Optimiz.}}\!\!\\
\cline{2-3}
& \!\!\!\! {Computation}\!\!\!  & \!\!\! {Memory}\!\!   &  \!\!\!\textbf{Free}\!\! & \!\textbf{Support} & \!\!\textbf{Level} \\
\hline
\!\!\!\textbf{ELSA} \cite{ham2021elsa} & \checked & $\times$ & $\times$ & $\times$ & Value\\
\rowcolor{mygray}\!\!\!\textbf{Sanger} \cite{lu2021sanger} & \checked   & $\times$ & $\times$ & $\times$ & Value\\
\!\!\!\textbf{DOTA} \cite{qu2022dota} & \checked & $\times$ & $\times$ & $\times$ & Value\\
\rowcolor{mygray}\!\!\!\textbf{DTATrans}\cite{yang2022dtatrans}\!\!\! &  \checked   & Low & \checked* & $\times$ & Value\\
\!\!\!\textbf{SpAtten} \cite{wang2021spatten}  & \checked  & Low & \checked* & $\times$ & \!\!\!Multi-bit\!\!\! \\
\rowcolor{mygray}\!\!\!\textbf{Energon} \cite{zhou2022energon} & \checked   & $\times$ & $\times$ & $\times$ & \!\!\!Multi-bit\!\! \\
\!\!\!\textbf{FACT} \cite{qin2023fact}  & \checked  & $\times$ & $\times$ & $\times$ & Value\\
\rowcolor{mygray}\!\!\!\textbf{SOFA} \cite{wang2024sofa} & \checked   & Low & $\times$ & \checked & Value\\
\!\!\!\textbf{PADE} & {\fontsize{6.5}{6} \CheckmarkBold }  & {\fontsize{6.5}{6} \CheckmarkBold } & {\fontsize{6.5}{6} \CheckmarkBold }  & {\fontsize{6.5}{6} \CheckmarkBold } & \textbf{Bit} \\
\specialrule{0.12em}{0.5pt}{0.1pt}
\end{tabular}
\begin{tablenotes}
\footnotesize
\item \hspace{-5mm}*  \!\!\! Sparsity guided by preceding layer scores; Accuracy degradation w/o retrain.
\end{tablenotes}
\end{threeparttable}
\label{tab:works_comparision}\vspace{-5mm}
\end{table}

\textbf{Key idea}. We reveal and leverage the monotonicity of the softmax function, and further adjust the pruning decision mechanism as follows: retaining tokens that reach the least significant bit (LSB) plane but without being pruned. This enables efficient and I/O-friendly pruning within tiled regions.

\textbf{Unfortunately, current attention accelerators still suffer from computation and memory access inefficiencies,  as they fail to exploit bit-grained opportunities to eliminate the high-overhead predictor.} Table \ref{tab:works_comparision} summarizes their features. The majority of existing works \cite{ham20203,ham2021elsa,lu2021sanger,qu2022dota,yang2022dtatrans} focus on accelerating attention by alleviating computation overhead. For example, ELSA \cite{ham2021elsa}, Sanger \cite{lu2021sanger}, DOTA \cite{qu2022dota}, FACT \cite{qin2023fact} adopt techniques like binary hashing, half-precision MSB, low-rank approximation, log-domain shifting to accelerate computation. However, these methods overlook memory optimization. While SpAtten \cite{wang2021spatten}. SOFA \cite{wang2024sofa} realizes this challenge, their coarse-grained strategies, like hybrid quantization, and cross-stage tiling, fail to exploit fine-grained, bit-level optimizations. Further, all current works rely on extra sparsity predictors, incurring substantial overhead. Notably, DTATrans \cite{yang2022dtatrans} and SpAtten \cite{wang2021spatten} guide sparsity using attention scores from the previous layer. While this strategy partially reduces predictor overhead, it necessitates resource-intensive retrain to recover accuracy. \textbf{These limitations motivate us to design an efficient attention accelerator that jointly optimizes computation and memory at fine granularity, while eliminating the sparsity predictor. }


\section{Algorithm Optimizations of PADE}
To effectively support the BSF strategy, we propose three key optimizations: BUI-GF, BS-OOE, and ISTA. BUI-GF ensures precise pruning in bit-wise operations, BS-OOE optimizes hardware utilization, and ISTA maintains sparsity while enhancing I/O efficiency via tiling attention.

\begin{figure}[t]
\centering
\includegraphics[width=\linewidth]{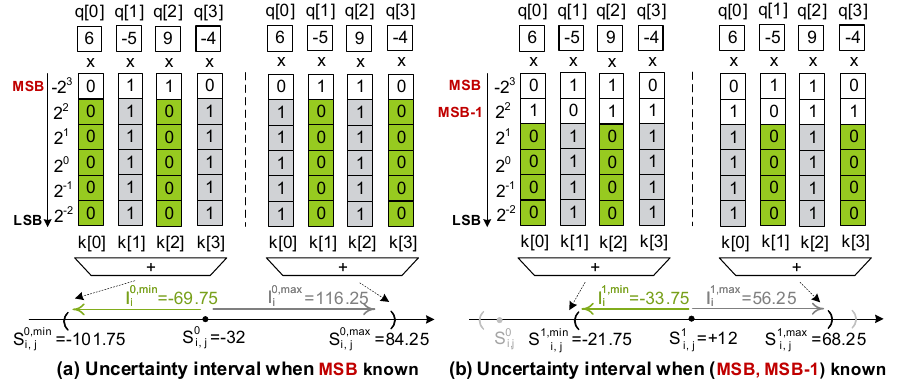}\vspace{-3mm}
\caption{Illustration of BUI with dot-product $\mathbf{Q}_i$ $\times$ $\mathbf{K}_j$.}
\label{fig:BUI}\vspace{-4mm}
\end{figure}

\subsection{BUI-enabled Guarded Filtering (BUI-GF) }\label{subsec:BUI-GF}
We begin by analyzing the mathematical properties of the softmax function, revealing its intrinsic potential for a simple yet efficient max-based pruning decision. Without loss of generality, consider a two-element input vector $[x_0,x_1]$, where $x_1$ is the max element and $x_1$$=$$x_0$$+$$\Delta$. As depicted in Eq.\eqref{eq:softmax}, the softmax output for $x_0$ decays exponentially with the gap $\Delta$ from $x_1$. This indicates that an element's contribution diminishes rapidly as its distance $\Delta$ from the max increases.
\begin{equation}
\!\!\!x_1\!\!=\!\! x_0 \!+\! \Delta \! \Rightarrow \! \mathrm{softmax}(x_0)\! = \! \frac{e^{x_0}}{e^{x_0}\! + \!e^{x_0 + \Delta}} \!=\! \frac{1}{1\! +\! e^{\Delta}}<\frac{1}{e^{\Delta}}\!\!
\label{eq:softmax}
\end{equation}

Naturally, a straightforward way to combine BSF with a max-based pruning strategy is to estimate the attention via partial bit planes of Keys, identify the max value, then use it to define a pruning threshold. However, such a crude method will incur severe estimation error, as analyzed in \S\ref{subsec:Challenges}.

To this end, we propose a \emph{bit-level uncertainty interval (BUI)} enabled guarded filtering (BUI-GF). We first introduce the BUI, which quantifies the potential variation in a dot-product, i.e., $\mathbf{Q}_i\mathbf{K}_j$ caused by the remaining bit planes of $\mathbf{K}_j$. Specifically, for a $p$-bit integer $b_{p-1}b_{p-2}...b_0$ with 2's complement format, its value $x$ is:     
\begin{equation}
x=-b_{p-1}2^{p-1}+\sum\nolimits_{i=0}^{p-2}b_i2^i.
\end{equation} 

\noindent In this format, all bits except the sign bit ($b_{p-1}$) contribute a non-negative value, meaning that each additional bit can only increase or maintain the magnitude. Based on this property, Figs.~\ref{fig:BUI} (a)(b) exemplify the BUI. In the example, $\mathbf{Q}_i$ denotes the $i$-th row of $\mathbf{Q}$ matrix, containing four entries with full 8-bit precision, while $\mathbf{K}_j$ is bit-serially processed: 1\,bit in Fig.~\ref{fig:BUI} (a) and 2\,bits in Fig.~\ref{fig:BUI} (b). For positive elements of $\mathbf{Q}_i$, BUI sets the unknown bits of $\mathbf{K}_j$ to 1 (shown in blue) and for negative entries, it set them to 0 (shown in red), yielding the potential largest score $S_{i,j}^{r,\max}$, as they account only for positive contributions. Here, $r$$\in$$[0,7]$ denotes the number of processed bit planes in 8-bit quantization. Conversely, BUI flips the unknown bits to obtain the potentially smallest score $S_{i,j}^{r,\min}$. We model this process as follows:
\begin{equation}
S_{i,j}^{r,\min}=S_{i,j}^r + I_i^{r,\min}, \hspace{4mm} S_{i,j}^{r,\max}=S_{i,j}^r + I_i^{r,\max},
\label{Eq:bit-margin}
\end{equation}

\noindent where $S_{i,j}^r$ denotes the conservative value by setting all unknown bits of $\mathbf{K}_j$ to zero, and $I_i^{r,\min}$ and $I_i^{r,\min}$ are the uncertainty intervals decided only by $\mathbf{Q}_i$. For example, in Fig.~\ref{fig:BUI} (a), we only know the MSB of $\mathbf{K}_j$, and by applying Eq. \eqref{Eq:bit-margin}, the BUI is determined to be the low bound (LB) $S_{i,j}^{0,\min}=-101.75$ and the upper bound (UB) $S_{i,j}^{0,\max}=84.25$. 

\begin{figure}[t]
\centering
\includegraphics[width=\linewidth]{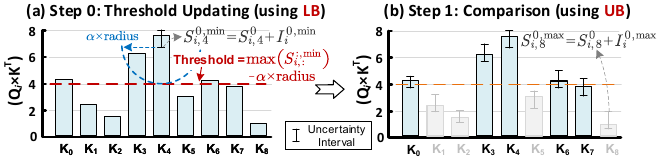}\vspace{-2mm}
\caption{Illustration of the BUI-GF strategy.}
\label{fig:Uncertainty_interval}\vspace{-4mm}
\end{figure}

Based on the BUI, we propose the BUI-GF strategy, which consists of two main steps, as illustrated in Fig. \ref{fig:Uncertainty_interval}. First, the BUI-GF determines the pruning threshold using the Eq. \eqref{eq:Threshold}:  
\begin{equation}
\mathcal{T}=\max(S_{i,:}^{:,\min})-\alpha \times radius, \hspace{2mm} 0\leq \alpha\leq 1,
\label{eq:Threshold}
\end{equation}

\noindent where $S_{i,:}^{:,\min}$ denotes the LBs of the estimated attention score for the $i$-th row. Notably, the scores are not confined to a specific bit plane, but are derived from all current processed bit planes. Based on our experiments, we set the default \emph{radius} to 5 and use a parameter $\alpha$$\in$$[0, 1]$ to control the threshold $\mathcal{T}$. By adjusting the $\alpha$, we can control the pruning ratio (DSE see \S\ref{subsec:Architecture_Evaluation}).
Then, the BUI-GF compares the UBs of attention scores with the threshold, retaining the Ks (e.g. 0,3,4,6,7), whose attention values are greater than this threshold. 

\subsection{BS-OOE: Improving Hardware Utilization}\label{subsec:BS-OOE}
The BUI-GF strategy enables accurate pruning decisions based on partial scores, which are incrementally estimated using bit-wise planes of the Key tensor. However, as analyzed in \S\ref{subsec:Challenges}, bit-grained execution will lead to hardware underutilization. To this end, BS-OOE employs a two-pronged approach: First, a bidirectional sparsity (BS) ensures load balancing across PEs. Second, bit-wise out-of-order execution (OOE) hides DRAM access latency. 


We extend the BS from static weight scenarios \cite{chen2024bbs} to dynamic attention workloads. The core idea of BS is that bit value `1' is also a form of sparsity. Without loss of generality, the dot-product between an $N$-element Query and Key can be formulated as:
\begin{equation}
\sum\nolimits_{j=0}^{N-1}q_{j}k_{j}=\sum\nolimits_{b=0}^{p-1}2^{b}\times\sum\nolimits_{j=0}^{N-1}q_j\times k_j^b,
\label{eq:Dual_sided1}
\end{equation}

\noindent where $k_j^b$ is the $b$-{th} bit of element $k_j$. Since each bit of $k_j$ can only be either 0 or 1, the second partial sum on the right-hand side of Eq. \eqref{eq:Dual_sided1} can be reorganized as:
\begin{equation}
\sum\nolimits_{j=0}^{N-1}\!\!q_j\times k_j^b =\! \!\!\sum\nolimits_{\forall j:k_j^b = 1}\! q_j = \!\!\sum\nolimits_{j=0}^{N-1}\!\!\! q_j \!-\!\! \sum\nolimits_{\forall j:k_j^b = 0} q_j.
\label{eq:Dual_sided2}
\end{equation}

From Eqs. \eqref{eq:Dual_sided1} \eqref{eq:Dual_sided2}, we observe that instead of accumulating the Query entries corresponding to bit-1, one can equivalently subtract the Query entries corresponding to bit-0 from the total sum of all Query entries. This transformation ensures that at most $50\%$ of the bits are involved in computation, thus effectively improving the load balance across PEs. 

Notably, unlike prior work \cite{chen2024bbs} that targets static weight compression or bit skipping, we extend bidirectional sparsity as a runtime load-balancing mechanism to bound PE imbalance in bit-serial QK execution, where a lightweight sparsity-aware scheduler (see \S\ref{subsec:ANDer_Tree}) is crucial due to the highly dynamic, runtime-determined nature of attention workloads.

\begin{figure}[t]
\centering
\includegraphics[width=\linewidth]{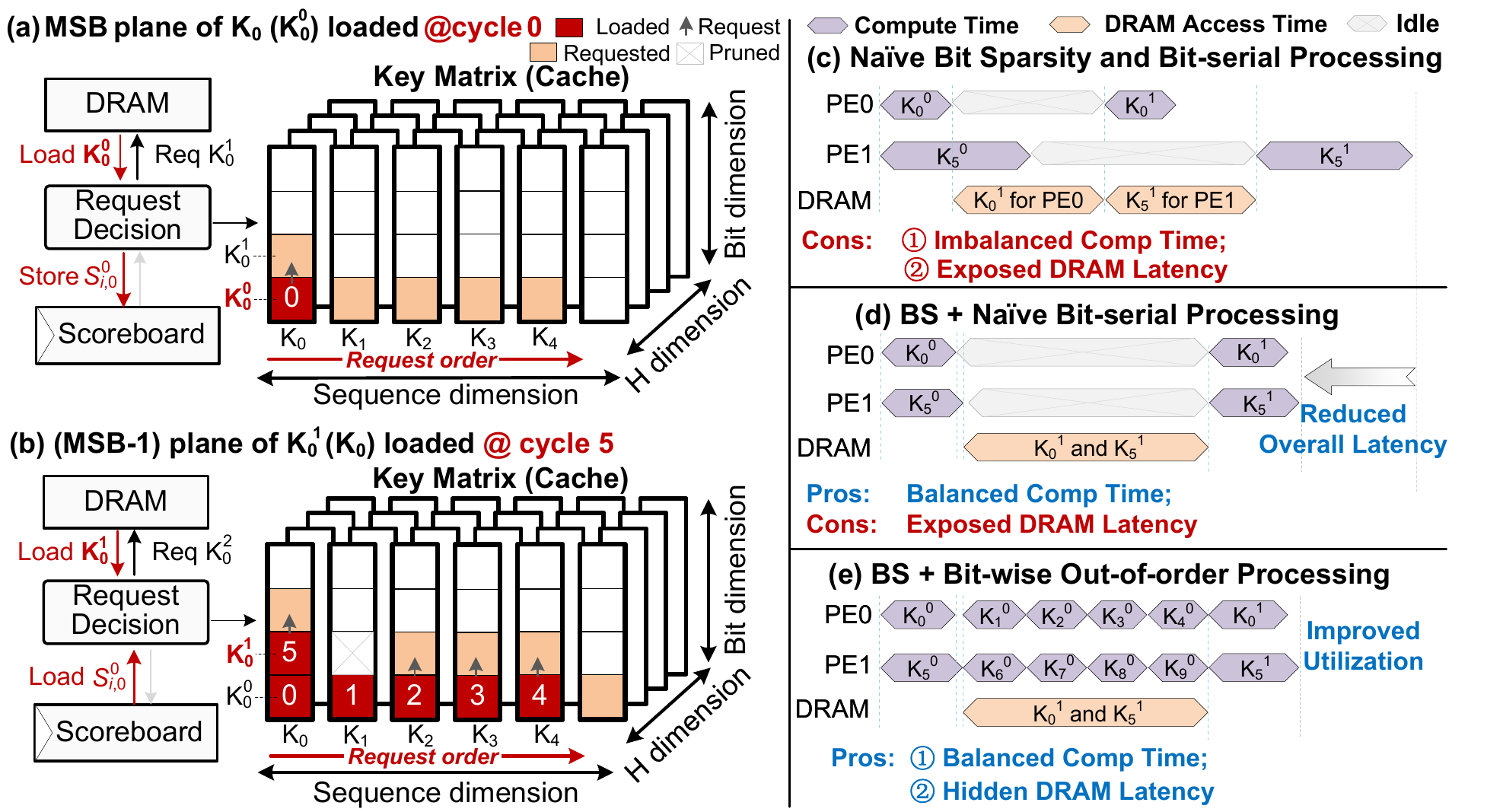}\vspace{-2mm}
\caption{(a)(b) Bit-wise OOE execution, (c)-(e) benefits of BS-OOE strategy.}
\label{fig:BOOE}\vspace{-5mm}
\end{figure}

Collaborating with BS, we propose the OOE strategy, as depicted in Fig. \ref{fig:BOOE} (a)(b), which operates as follows: \ding{182} When the score speculation begins, only the first bit planes of Key vectors are requested. \ding{183} Once any bit plane is loaded from DRAM, its partial score is immediately computed, followed by the BUI-GF pruning decision. \ding{184} If not pruned, the next bit plane of that Key vector is requested (e.g. the requested K$_0^1$ in Fig. \ref{fig:BOOE} (a)), while its partial score ($S_{i,0}^0$) is stored in a Scoreboard. Otherwise, the process proceeds by requesting the first bit plane of the next Key vector. Before the required bit plane is loaded on chip, the PE continue processing other Keys, such as $K_1^0$,..,$K_4^0$. \ding{185} When the downstream bit plane (e.g., $K_0^1$ in Fig. \ref{fig:BOOE} (b)) is loaded from DRAM, it retrieves the corresponding partial score (i.e., $S_{i,0}^0$) from the Scoreboard and updates it with the newly computed partial score. The updated score is then used to repeat steps \ding{184} and \ding{185}. In this way, the compute units remain active, thereby improving hardware utilization for bit-wise speculation. 

Figs. \ref{fig:BOOE} (c)-(e) highlight the advantages of BS-OOE over naive bit sparsity + bit serial processing, by comparing the timeline of PE0 in the scenarios depicted in Figs. \ref{fig:BOOE} (a)(b). As shown in Fig. \ref{fig:BOOE} (c), naive bit sparsity processing suffers from workload imbalance, causing severe computation time discrepancies across PEs. This results in scattered DRAM accesses, idle cycles, and high latency. In Fig. \ref{fig:BOOE} (d), BS alleviates this imbalance between PEs, enabling memory request merging and reducing row activation during DRAM access. However, computation resources remain underutilized during memory access. In contrast, as depicted in Fig. \ref{fig:BOOE} (e), BS-OOE further leverages bit-wise out-of-order execution to improve resource utilization. Specifically, while PE0 waits for DRAM to return $K_0^1$, it continues to process $K_1^0,..\,, K_4^0$, without idling.

\subsection{ISTA: Enhancing IO Efficiency}\label{subsec:TermiFlashAttention} 
Reducing unnecessary data movement is critical for supporting ultra-long sequences. While tiling (e.g., FlashAttention) offers a promising solution, it is incompatible with sparse attention because it breaks the row-wise dependency of softmax, on which BUI-GF (\S\ref{subsec:BUI-GF}) relies for max-based pruning.


\begin{figure}[b]
\vspace{-4mm}
\centering
\includegraphics[width=\linewidth]{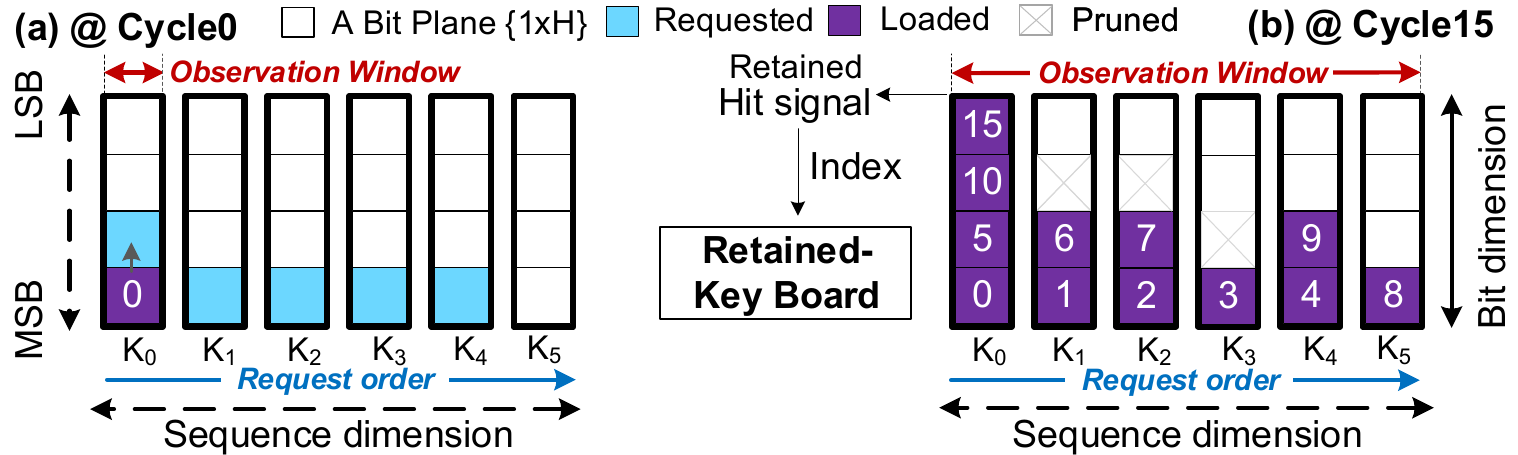}\vspace{-3mm}
\caption{Tile-level sparsity decision.}
\label{fig:Tiling_Decision}\vspace{-0mm}
\end{figure}

To address this, we re-examine softmax properties and show that pruning decisions can be performed on subsets with proper adjustments. Eq. \eqref{eq:softmax_mon} shows that the softmax denominator grows monotonically as more elements are added, due to the non-negativity of exponentials. Therefore, if a token's score falls below the threshold within a subset, its global row-wise softmax score can only be lower. This enables the safe application of the BUI-GF strategy within tiled regions.
\begin{equation}
softmax(x_i)=\frac{\exp(x_i)}{\sum_{j=0}^{N-1}\exp(x_j)}\leq\frac{\exp(x_i)}{\sum_{j\in subset} \exp(x_j)}
\label{eq:softmax_mon}
\end{equation}

\begin{figure}[t]
\centering
\includegraphics[width=\linewidth]{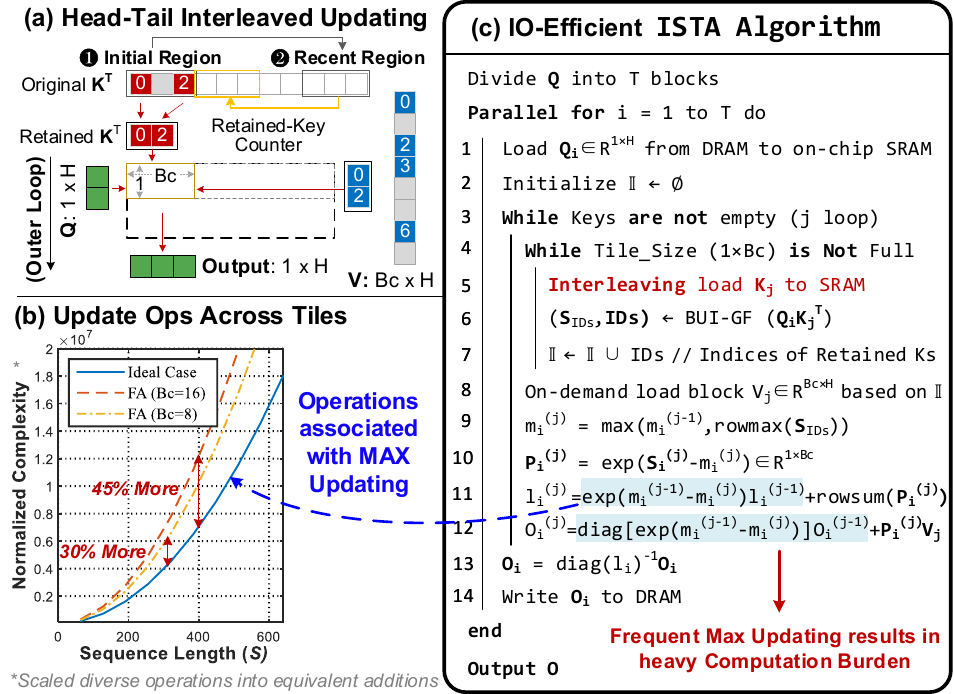}\vspace{-2mm}
\caption{(a) Illustration for head-tail interleaving updating. (b) Heavy updating cost across tiles. (c) Detailed procedure of the ISTA algorithm.}
\label{fig:TermiFlashAttention}\vspace{-5mm}
\end{figure}

To identify retained keys within a subset, a key is retained if all its bit planes are processed and it remains unpruned. As depicted in Fig. \ref{fig:Tiling_Decision}, at cycle 0, only the partial score of $K_0$ is calculated, with the subset observation window size initialized to 1. By cycle 15, scores relative to six Keys have been computed, expanding the window to size 6. During each bit plane calculation, the BUI-GF works continuously to check if pruning. If a Key (i.e., $K_0$) remains unpruned after processing its least significant bit (LSB) plane, it is deemed an important token, and its index is stored in the Retained Key Board.

\begin{figure*}[t]
\centering
\includegraphics[width=\linewidth]{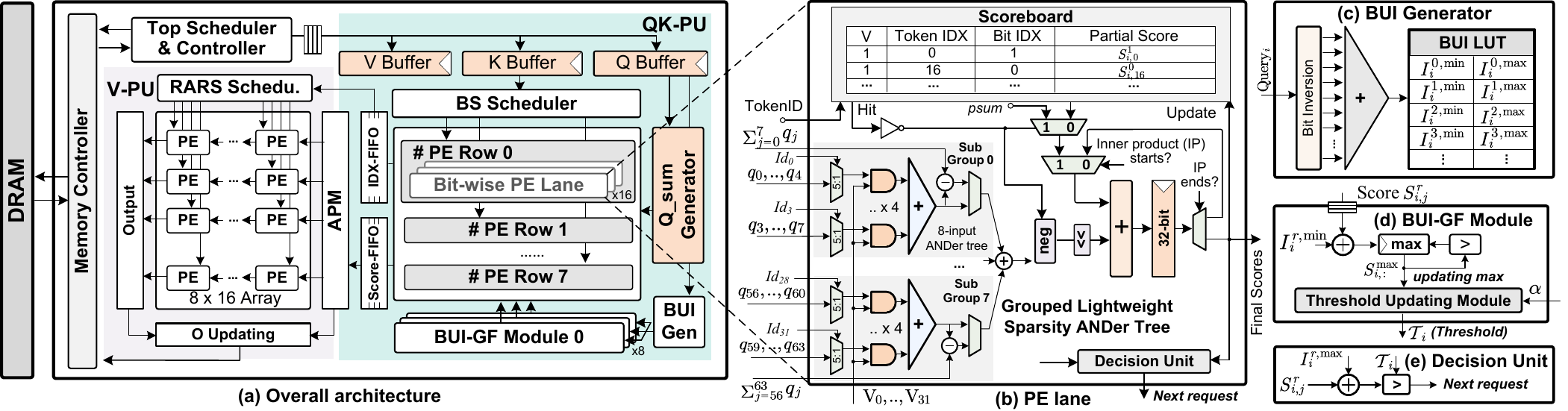}\vspace{-2mm}
\caption{(a) Overview of the PADE architecture. (b) Bit-wise PE lane. (c) BUI Generator. (d) BUI-GF Module. (e) Decision Unit.}
\label{fig:Overall_architecture}\vspace{-3mm}
\end{figure*}

Driven by the tile-level sparsity decisions, the ISTA algorithm is detailed in Fig. \ref{fig:TermiFlashAttention} (c). Its process begins by performing sparse QK computations, retaining and storing the indices (IDs) and scores (S$_{\rm IDs}$) of selected Keys (lines\,4-6). Once the number of retained entries reaches the tile size $B_c$ (line\,3), the corresponding Vs are fetched for subsequent computation (line\,8). As more tiles are processed, the maximum value (line\,9) and the exponential sum (line\,11) are progressively updated and propagated across tiles. Finally, the attention output is computed based on the accumulated results (line\,13).

However, a key observation is that naive left-to-right computation causes frequent updates of the maximum across tiles (line 9 in Fig. \ref{fig:TermiFlashAttention} (c)), leading to a series of redundant operations (line 11). Specifically, each time the maximum is updated, it triggers one subtraction, one exponentiation and two multiplications of scalar and vector (see highlights in lines 11-12). We employ the arithmetic complexity model \cite{brent2010modern,wang2021efficientMIMO,wang2023efficient} to normalize the complexity for different operations. As profiled in Fig. \ref{fig:TermiFlashAttention} (b), for $S$=$2048$ and $B_c$=16, this leads to about 30\% increase in computation complexity compared to the vanilla implementation. Moreover, the overhead becomes more as $B_c$ decreases, due to more frequent updates.


To minimize the overhead associated with the `max updating', it is advantageous to prioritize dominant tokens. Note that the reduction in operations refers to the redundant computations shown in lines 11–12, rather than the comparison operations in line 9. To this end, we propose a \textit{head-tail interleaved updating} strategy that exploits the locality property of attention: Recently generated tokens and the initial token typically exhibit higher weights than others \cite{tian2023scan,jiang2024minference}. As illustrated in Fig. \ref{fig:TermiFlashAttention}(a), the update begins with the initial region, then jumps to the recent region, and subsequently returns to the post-initial region, repeating this interleaved pattern across tiles. This pattern reduces unnecessary maximum value updates, resulting in a $20\%$-$40\%$ reduction in total operations. It is important to note that, without attention locality, the performance of head-tail interleaving is on par with regular execution and not worse.

\section{Architecture and Hardware Innovation}\label{sec:Architecture} 
\subsection{Architecture Overview} 
Fig. \ref{fig:Overall_architecture} (a) depicts the overall architecture of the PADE, which incorporates two major components:

1) Query-Key Processing Unit (QK-PU): This unit computes the dot-product of the Query and Key matrices (i.e., $\mathbf{QK}^T$). It includes a PE array, eight BUI-GF modules, a Q$\_$sum generator, a sparsity scheduler, and a BUI generator. The PE array comprises eight rows, each containing 16 bit-wise PE lanes, dedicated to processing a single query. Together, these components support both the BUI-GF and BS-OOE strategies.

2) Value Processing Unit (V-PU): To support ISTA, the VPU unit computes final results from retained (non-pruned) scores in a tiled manner. It comprises an auxiliary processing module (APM) for exponentiation, followed by an 8$\times$16 output-stationary systolic array. To reduce memory access overhead, a RARS scheduler is integrated to enhance data reuse.


\subsection{Overall Dataflow} 
In PADE, self-attention operands use 8-bit precision, with each Key vector divided into eight 1-bit planes. During the prefill stage, PADE processes eight queries within a head, whereas during the decoding stage, it handles different queries across multiple heads. Its detailed process is as follows (Fig.~\ref{fig:Overall_architecture} (a)):


First (\ding{182}), before the $\mathbf{Q}_i\mathbf{K}^T$ computation, the BUI Generator initializes eight uncertainty interval pairs $(I_{i}^{r,\min}, I_{i}^{r,\max})$, where $r$\,$\in$\,$[0,7]$, based on the input $\mathbf{Q}_i$. Each pair corresponds to a specific bit plane. These uncertainty interval pairs are then stored in a lookup table (LUT), as shown in Fig. \ref{fig:Overall_architecture} (c). Next (\ding{183}), $16$ PE lanes in a PE row perform the dot product for $\mathbf{Q}_i\mathbf{K}^T$ and in an out-of-order manner in parallel. Following this (\ding{184}), the BUI-GF Module, as depicted in Fig. \ref{fig:Overall_architecture} (d), calculates the pruning threshold $\mathcal{T}_i$ using the max value among the all current score $S_i^{:,\min}$ with the BUI-GF logic (Eq. \eqref{eq:Threshold}). Finally (\ding{185}), the pruning threshold $\mathcal{T}_i$ is broadcast to all PE lanes in a row, enabling the evaluation of whether the score of the token $j$ satisfying $S_{i,j}^{r,\max}$$>$$\mathcal{T}_i$, as depicted in Fig. \ref{fig:Overall_architecture} (e). If true, the PE lane requests the next bit plane for further computation. Otherwise, the token $j$ is immediately pruned. This process is repeated until the LSB is reached. The remaining scores are sent to V-PU to produce final outputs. 

\subsection{ScoreBoard-Based Result Reusable PE Lane}\label{subsec:ScoreBoard-Based Result Reusable PE Lane}
To perform bit-serial speculation for $\mathbf{QK}$, a straightforward solution is to incrementally compute over the bit planes of $\mathbf{K}$. Specifically, it accesses one bit plane (MSB) in the first round, two bit planes (MSB, MSB-1) in the second round, and so on, until the final round. However, repeated memory accesses in this scheme lead to significant power consumption. 

To enable result reuse, we dedicate a scoreboard-assisted bit-wise PE lane, as shown in Fig. \ref{fig:Overall_architecture} (b), which comprises three key components: a grouped lightweight sparsity ANDer tree (GSAT) for 64-input dot-product and two modules supporting BUI-GF for out-of-order execution. 1) To avoid repeatedly loading bit planes, each PE lane integrates a scoreboard that temporarily caches partial scores $S_{i,j}^r$ for unpruned tokens. 2) The Decision Unit performs BUI-GF logic, makes pruning decisions and selects the next bit plane to fetch. 

These modules operate collaboratively to enable efficient early pruning while minimizing redundant memory access. Initially, the GSAT computes the partial dot product $\Delta S_{i,j}^r$ from an 8-bit vector $\mathbf{Q}_i$ and a 1-bit Key plane $\mathbf{K}_j^r$. Meanwhile, the scoreboard is accessed via the Key's index $j$. If a previous partial score $S_{i,j}^{r-1}$ exists, it is fetched  and updated as $S_{i,j}^r$$=$$S_{i,j}^{r-1}$ $+$ $\Delta S_{i,j}^r$. Otherwise, indicated that the current bit plane is MSB, $\Delta S_{i,j}^r$ is directly written to the scoreboard, and Hit signal is pulled down to show no prior score is available.

\emph{Decision Unit.} The unit determines pruning by receiving the max uncertainty interval $I_{i}^{r,\max}$ and the partial score $S_{i,j}^r$. It checks whether $S_{i,j}^r$$+$$I_{i}^{r,\max}$$>$$\mathcal{T}_i$ holds. If true, it requests the next bit plane of $\mathbf{K}_j$, i.e., $\mathbf{K}_j^{r+1}$, and updates the partial score in the scoreboard. Otherwise, it evicts the token entry from the scoreboard and requests the next Key vector from DRAM.

\emph{BUI-GF Module.} As shown in Fig. \ref{fig:Overall_architecture} (d), the BUI-GF module reads scores from registers and adds the min uncertainty interval $I_i^{r,\min}$ to compute the lower bound of scores. Based on these values and a predefined ratio $\alpha$, the Threshold Updating Module applies BUI-GF logic (Eq. \eqref{eq:Threshold}) to generate threshold $\mathcal{T}_i$, which is then broadcast to all PE lanes in row$_i$.

\subsection{Grouped Lightweight Sparsity ANDer Tree (GSAT)}\label{subsec:ANDer_Tree}
\begin{figure}[t]
\centering
\includegraphics[width=\linewidth]{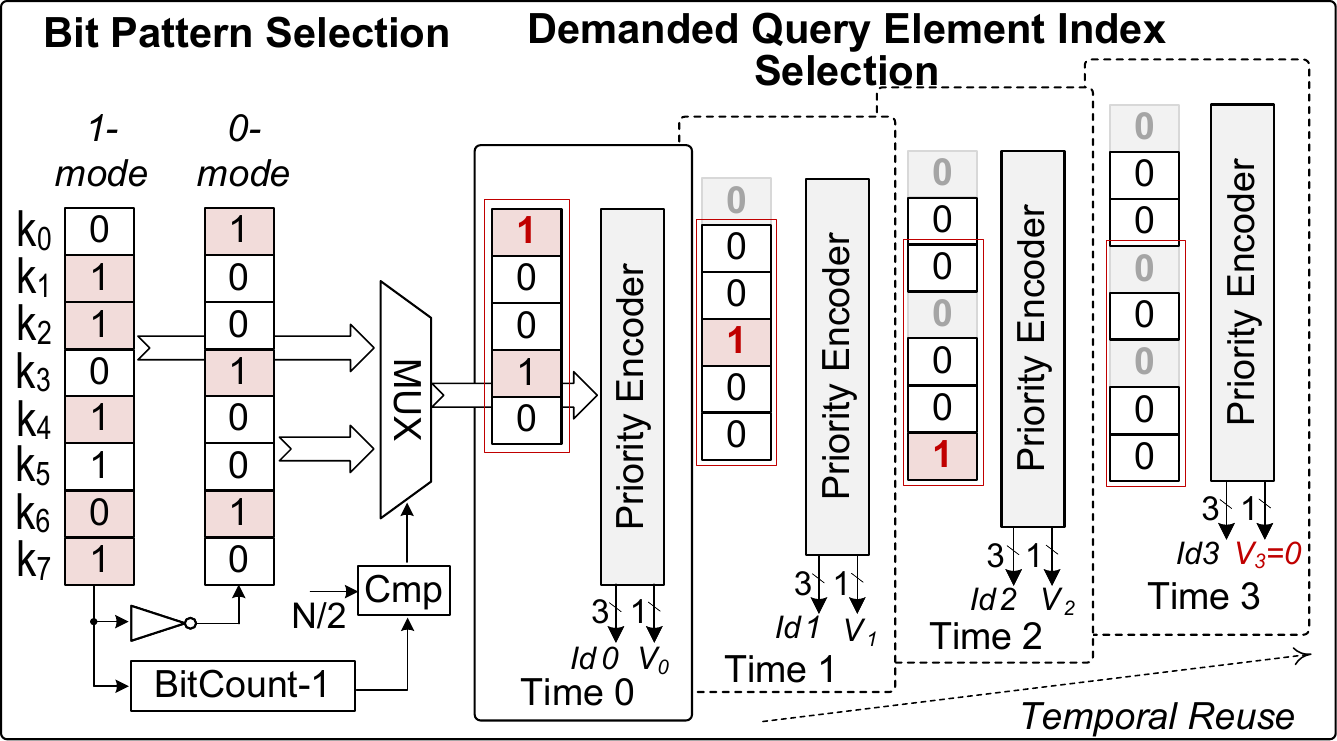}\vspace{-2mm}
\caption{Temporal reuse architecture of bidirectional sparsity (BS) scheduler.}
\label{fig:BS_Scheduler}\vspace{-3mm}
\end{figure}
To exploit bit sparsity, a direct design is to select and accumulate activations corresponding to non-zero bits. However, this requires large multiplexers, reducing efficiency. Specifically, to perform a 64-dimensional dot product, such a coarse design will require at least 32 64-input multiplexers.




Given BS guarantees at least $50\%$ sparsity in a bit-vector of arbitrary length, it is possible to reduce the MUX cost with a smaller group size. Based on this insight, we propose a grouped lightweight ANDer tree architecture. As depicted in Fig. \ref{fig:Overall_architecture} (b), for a 64-input dot-product, we first decompose it into eight accumulation sub-groups, each comprising 8 dimensions. For each sub-group, in the worst-case scenario, the selected query elements within the sub-group $\{q_0,\cdots,q_7\}$ will be $\{q_4,\cdots,q_7\}$. Therefore, in a sub group, only four 5:1 multiplexers are required: the first multiplexer selects among $\{q_0,\cdots,q_4\}$, the second among $\{q_1,\cdots,q_5\}$, and so on. Compared to the naive design, MUX cost is reduced (32×64:1 $\to$ 32×5:1), with a trade-off of more subtractors.      

\textbf{Insight: There is an optimal sub-group size that minimizes hardware overhead.} A smaller sub-group size reduces multiplexer overhead but adds extra cost from subtractors and Q\_sum generators. The DSE is provided in Fig. \ref{fig:hardware-trade-off} (a).  

\textbf{Lightweight BS Scheduler}. Built upon \cite{chen2024bbs}, PADE adopts a low-cost BS scheduler by reusing a critical priority encoder across time steps to orchestrate operations within each PE, as depicted in Fig. \mbox{\ref{fig:BS_Scheduler}}. Unlike the straightforward design in \cite{chen2024bbs}, which instantiates multiple priority encoders in parallel, PADE temporally multiplexes a single priority encoder across successive time steps for index selection. This temporal reuse is enabled by PADE’s distinctive pipelined microarchitecture, where the QK-PU and V-PU operate in a staggered pipeline, effectively hiding the additional latency introduced by encoder reuse. The temporal reuse scheme helps PADE reduce 75\% priority encoder overhead.

To control the bit-serial dot product, the scheduler first identifies whether a bit plane of the Key vector contains more zeros or ones. It then sends the original or flipped bit column to a priority encoder. The priority encoder operates on five consecutive bits of the plane at a time. For example, at the first time, it receives \{$k_0$,..,$k_4$\}, followed by \{$k_1$,..,$k_5$\} in the next, and so on. The encoder detects the location of the first ``1" bit within each 5-bit vector. If such a bit exists, it is masked, and the remaining bits are propagated to the next time step. Otherwise, if the vector contains only zeros, the encoder asserts \texttt{V}$_x$\texttt{=0} to disable the corresponding bit-serial multiplier in the PE, as depicted in Fig. \mbox{\ref{fig:Overall_architecture}} (b).

\begin{figure}[t]
\centering
\includegraphics[width=0.96\linewidth]{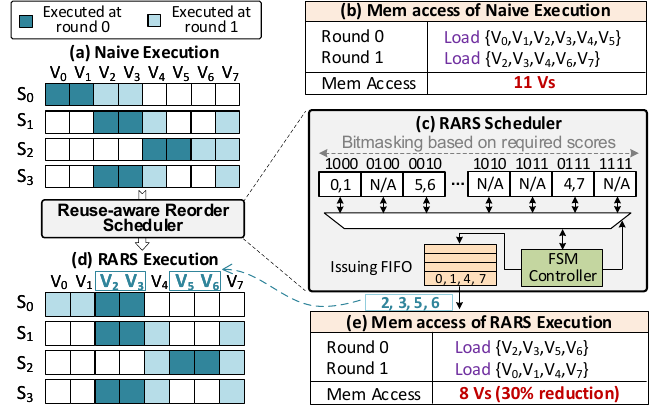}\vspace{-2mm}
\caption{Reuse-aware reorder scheduler (RARS).}
\label{fig:RARS}\vspace{-5mm}
\end{figure}

\begin{table*}[t]
\scriptsize
\renewcommand{\arraystretch}{0.95}
\caption{Accuracy of Different Transformer Models with MXINT8, FP16, INT8 and PADE Configurations (S: Standard, A: Aggressive).}\vspace{-3mm}
\begin{tabular}{l|m{0.3cm}<{\centering}m{0.3cm}<{\centering}m{0.3cm}<{\centering}m{0.35cm}<{\centering}m{0.35cm}<{\centering}m{0.4cm}<{\centering}m{0.3cm}<{\centering}m{0.35cm}<{\centering}m{0.35cm}<{\centering}m{0.35cm}<{\centering}m{0.35cm}<{\centering}m{0.35cm}<{\centering}m{0.35cm}<{\centering}m{0.3cm}<{\centering}m{0.3cm}<{\centering}m{0.3cm}<{\centering}m{0.3cm}<{\centering}m{0.3cm}<{\centering}m{0.3cm}<{\centering}m{0.3cm}<{\centering}m{0.25cm}<{\centering}c}
\specialrule{0.12em}{0.5pt}{0.8pt}
\!\!\textbf{Model} & \multicolumn{6}{c}{LlaMa2-7B} & \multicolumn{6}{c}{LlaMa3-8B} & \multicolumn{2}{c}{OPT1B3} & \multicolumn{2}{c}{Bloom1B7} & \multicolumn{2}{c}{Qwen7B} & \multicolumn{2}{c}{ViT-L/16} & \multicolumn{2}{c}{PVT} \\
\hline
\!\!\textbf{Task}$^{\ddagger}$ &  \!\!Dolly  & \!\!\!\!Wikili. & \!\!\!MBPP & \!\!Wiki2 &  \!\!\!MMLU\!\!\!\! & \!\!Winog. & \!\!Dolly   & \!\!Wikili. & \!MBPP & \!Wiki2 & \!\!MMLU & \!\!Winog.   & \!\!Wikili. & \!\!MBPP & \!\!Wikili. & \!\!MBPP & \!\!Wikili. & \!\!MBPP & \!Image & \!\!VTAB & \!\!Image & VTAB\\
\hline
\!\!\textbf{MXINT8} & \!\!36.5  &  \!39.3 & \!\!17.5\% & 5.63 & \!\!35.2\% & \!\!69.8\%  & \!40.9 & \!43.6 & 23.3\% & 5.01 & 42.2\% & 75.1\% & 36.1 & 11.9\% & 44.6 & 16.3\% & 46.8 & 30.5\% & 85.5\% & 72.8\% & \!89.7\% & 77.5\%\\

\!\!\textbf{FP16} & \!\!36.4  &  \!39.1 & \!\!17.5\% & 5.71 & \!\!35.1\% & \!\!69.4\%  & \!40.8 & \!42.7 & 21.8\% & 5.11 & 41.2\% & 74.2\% &    36.2 & 11.9\% & 44.3 & 16\% & 46.6 & 30\%  & 85.3\% & 72.7\% & \!89.4\% & 77.3\%\\

\!\!\textbf{INT8} & \!\!36.4  & \!38.9 & \!\!17.2\% & 5.73 & \!\!34.7\% & \!\!69.3\%  & \!40.7 & \!42.7 & 21.6\%  & 5.13 & 40.9\% &  73.7\% &  35.9 & 11.6\% &  44.1 & 15.7\% & 46.4 & 29.2\% & 85.3\% & 72.5\% & \!89.3\% & 77.1\% \\

\!\!\textbf{PADE (S)}\!\! & \!\!36.3  & \!38.9  & \!\!17.2\% & 5.75 & \!\!34.6\% & \!\!69.2\%  & \!40.6 & \!42.6 & 21.5\%  & 5.13 & 40.7\% & 73.7\% & 35.9 & 11.5\% & 44.0 & 15.6\% & 46.3 & 29.2\% & 85.3\% & 72.5\% & \!89.3\% & 77.1\% \\

\!\!\textbf{PADE (A)}\!\! & \!\!36.1  & \!38.4 & \!\!16.5\% & 5.80 & \!\!34.1\% & \!\!68.7\%  & \!40.5 & \!42.0 & 21.0\%  & 5.19 & 40.2\% & 72.8\% &  35.3 & 11.0\% &43.6 & 15.2\% & 45.9 & 28.4\% & 84.9\% & 72.4\% & \!89.1\%  & 76.8\%\\
\specialrule{0.12em}{0.5pt}{0.5pt}
\end{tabular}\vspace{0mm}
\begin{footnotetext}
\scriptsize
$^{\ddagger}$ MMLU, WinoGrande, MBPP, Imagenet, VTAB are evaluated by accuracy. Dolly, Wikilingua are evaluated by ROUGE-1. Wikitext2 is evaluated by PPL, where lower is better.
\end{footnotetext}
\label{tab:accuracy}\vspace{-1mm}
\end{table*}

\begin{figure*}[t]
\centering
\includegraphics[width=\linewidth]{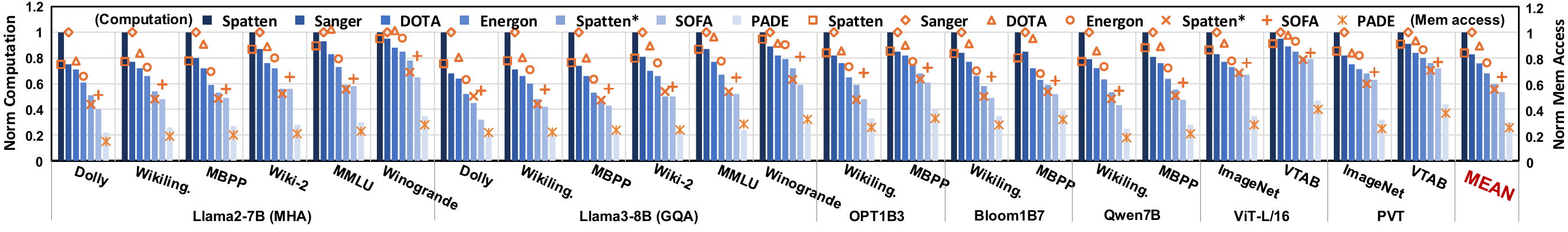}\vspace{-3mm}
\caption{Normalized computation and memory access across diverse Transformer models and tasks. Spatten* performs additional fine-tuning.}
\label{fig:Overall_Complexity}\vspace{-4mm}
\end{figure*}

\subsection{Reuse-Aware Reorder Scheduling (RARS)}\label{subsec:RARS}
Due to attention sparsity, the remaining scores are distributed irregularly. For example, in Fig. \ref{fig:RARS} (a), the 0-th score row S$_0$ retains elements at positions 0-3, which are multiplied with the corresponding V$_0$-V$_3$ vectors (refer to Fig. \ref{fig:attention_mechanism}).  

However, a naive design leads to redundant V-vector memory accesses during the computation of $\mathbf{S} \times \mathbf{V}$. As shown in Fig. \ref{fig:RARS} (a), assuming each PE row in V-PU processes two V vectors per round, a naive left-to-right execution computes the dot products (S$_0$,V$_0$V$_1$), (S$_1$,V$_2$V$_3$), (S$_2$,V$_4$V$_5$), (S$_3$,V$_2$V$_3$) in round 0, followed by (S$_0$,V$_2$V$_3$), (S$_1$,V$_4$V$_7$), (S$_2$,V$_6$V$_7$), (S$_3$,V$_4$V$_7$) during round 1. Due to data reuse inefficiency, some shared V vectors must be reloaded, resulting in a total of 11 V-vector memory accesses.

To implement the ISTA strategy efficiently, we propose reuse-aware reorder scheduling (RARS) to reduce redundant memory access. As shown Fig. \ref{fig:RARS} (d), V$_2$ and V$_3$ are shared among three scores: S$_0$, S$_1$ and S$_3$, making them are prioritized for initial scheduling. Then, RARS selects V$_5$ and V$_6$, which are used exclusively by the remaining score S$_2$. As a result, V$_2$, V$_3$, V$_5$ and V$_6$ are grouped for execution in round 0. Such greedy search continues until all scores are allocated adequate Vs. As depicted in Fig. \ref{fig:RARS} (e), compared to the default left-to-right computation order, RARS reduces $30\%$ memory access.   


We design an efficient scheduler to implement RARS. As shown in Fig. \ref{fig:RARS} (c), condition statements and control logic are handled by an FSM controller. A single-port read-write ID buffer, indexed by score-derived bitmasks, stores the corresponding V-vector IDs. For example, V$_5$ and V$_6$, used exclusively by S$_2$, are stored in buffer-0010. Guided by RARS logic, the FSM retrieves the buffer entries and dispatches them to the issuing FIFO in an optimized execution order.

\section{Evaluation}\label{sec:Evaluation}
\subsection{Experimental Setup}\label{subsec:Experimental_setup}
\textbf{Baseline comparisons}: We compare PADE with five SOTA attention accelerators: Sanger \cite{lu2021sanger}, Spatten \cite{wang2021spatten}, Energon \cite{zhou2022energon}, DOTA \cite{qu2022dota}, SOFA \cite{wang2024sofa}. For fair comparison, all designs are normalized to a 28nm process and evaluated under identical conditions: PE arrays occupy the same area as PADE and work in 800\,MHz, on-chip SRAM is set to 352kB, and peak HBM bandwidth is 256 GB/s, with 4 pj/bit \cite{o2017fine}.

\begin{table}[t]
\renewcommand{\arraystretch}{0.8}
\centering
\vspace{-1mm}
\scriptsize
\caption{On-chip Hardware and Off-chip DRAM Configurations of PADE}\vspace{-2mm}
\begin{tabular}{p{1.95cm}|m{5.9cm}<{\centering}}
\specialrule{0.12em}{0.5pt}{0.9pt}
\!\!\multirow{2}{*}{\textbf{On-chip Buffer}} & 320KB SRAM for Key and Value buffers; \\
                        & 32KB Q buffer \\ \specialrule{0.05em}{0.5pt}{0.9pt}
\!\!\multirow{2}{*}{\textbf{QK-PU}} & 128 Bit-wise PE Lanes; A Q$\_$sum Generator;\\
& \!\!\!\!\!\!8 BUI-GF modules; A BUI Generator; A Sparsity Scheduler\!\!\!\!\!\\
\specialrule{0.05em}{0.5pt}{0.9pt}
\!\!\multirow{2}{*}{\hspace{1mm}\textbf{- Bit-wise PE Lane}} & \!\!64-dim $\times$ 8-bit $\times$ 1-bit Grouped Sparsity ANDer tree;\!\! \\
                 & 32 entry $\times$ 45-bit Scoreboard \\ \specialrule{0.05em}{0.5pt}{0.9pt}
\!\!\multirow{2}{*}{\textbf{V-PU}} & A Systolic Array with 8 $\times$ 16 INT8 PEs; \\
                 & A 128-input FP16 APM; A RARS Scheduler \\
\specialrule{0.05em}{0.5pt}{0.9pt}
\!\!\multirow{2}{*}{\textbf{Off-chip DRAM}} & HBM2; 16$\times$64-bit pseudo channels@2Gbps;\\
& Each channel provides 16GB/s; BL=4$\times$64b; t{RC}=50ns\\
\specialrule{0.12em}{0.5pt}{0pt}
\end{tabular}
\label{tab:PADE_Architecture}\vspace{-3mm}
\end{table}

\textbf{Benchmarks}: We evaluate PADE on several representative Transformer models across NLP and CV tasks. For NLP tasks, Llama2-7B \cite{touvron2023llama2}, Llama3-8B \cite{grattafiori2024llama}, Qwen7B \cite{bai2023qwen}, Bloom1B7 \cite{le2022bloom} and OPT1B3 \cite{zhang2022opt}, for six tasks.These tasks include language modeling (Wikitext-2 (S=2k) \cite{merity2016pointer}, Wikilingua (S=2k) \cite{ladhak-wiki-2020}, Winogrande (S=0.25k) \cite{sakaguchi2021winogrande}), language understanding (MMLU, S=0.5k) \cite{hendrycks2020measuring}, code generation MBPP (S=1k) \cite{austin2021program}, long
context processing dolly (S=15k) \cite{conover2023free}. For CV tasks, we choose ViT-L/16 (S=576) \cite{dosovitskiy2020image} and PVT (S=3k) \cite{wang2021pyramid} on ImageNet-1k \cite{deng2009imagenet} and JFT \cite{sun2017revisiting} classification.



\textbf{Quantization Accuracy}. All pre-trained models are sourced from Pytorch \cite{paszke2017automatic} and HuggingFace \cite{wolf2020transformers}. INT8 baselines derived via post-training quantization, where only the weights and activations (QKVs) are quantized to INT8, while non-linear operators (e.g., softmax) remain in FP16 precision. As shown in Table. \ref{tab:accuracy}, the INT8 baseline incurs less than a $1\%$ average accuracy drop from FP16, thus confirming its validity. 

\textbf{Hardware Evaluation}: Table \ref{tab:PADE_Architecture} lists the hardware configuration of PADE. We implement the RTL design for PADE and utilize Synopsys DC on TSMC 28nm CMOS technology to estimate the logic area and power. The power, area, and read/write bandwidth of on-chip SRAM buffers are estimated through CACTI \cite{muralimanohar2009cacti}. Off-chip HBM modeling involves simulating access patterns and row activation under various data layouts (Fig. \mbox{\ref{fig:Memory_Layout}}), capturing HBM’s burst behavior. We derive memory latency from Ramulator \mbox{\cite{kim2015ramulator}}, and estimate IO power following the methodology in~\cite{cavigelli2016origami,andri2016yodann,wang2017energy}. We extract each stage’s cycles by simulating the RTL with Verilator~\mbox{\cite{snyder2004verilator}}, and use a custom cycle-level simulator to evaluate performance.

\begin{figure}[t]
\centering
\includegraphics[width=\linewidth]{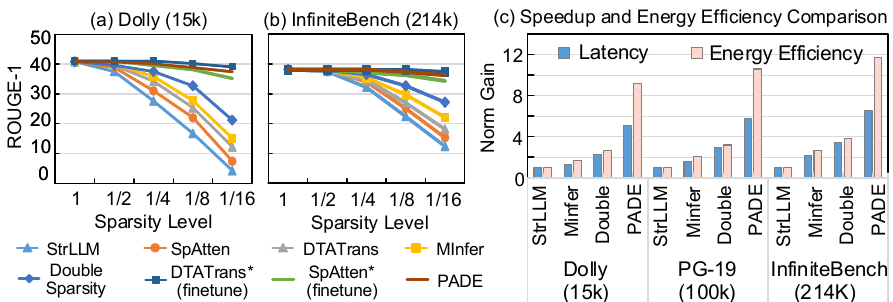}\vspace{-3mm}
\caption{(a)(b) Accuracy comparison with current sparse attention methods. (c) Speedup and efficiency gain comparison.}
\label{fig:SW-SparseAttention}\vspace{-3mm}
\end{figure}

\textbf{GPU comparison} We benchmark on an Nvidia H100 using SOTA TensorRT-LLM \cite{Nvidia2023} with FlashAttention3 \cite{shah2024flashattention}. \textbf{To exclude the software overhead}, we measure execution time with \emph{cudaEvent}, isolating GPU execution from CPU interference. The GPU is dedicated during testing, and large batch sizes are used to amortize data transfer costs. Non-computational phases are excluded using \emph{nvprof}. Power is measured via \emph{nvidia-smi}; dynamic power is computed as the difference between active and idle states. Each experiment is run 2k times, discarding the top and bottom $15\%$ before averaging. 

\textbf{GPU test configurations}: We test all different datasets with their allowed sequence lengths ranging from 0.25k to 15k. For example, MMLU (0.5k) and Dolly (15k). We measure the total inference latency, including the prefill and decoding. Specific prefill and decoding lengths are determined by the dataset itself. For batch size, we will select the configuration from [8, 128] that maximizes GPU computational efficiency based on the sequence length of each dataset.



\begin{figure}[t]
\centering
\includegraphics[width=\linewidth]{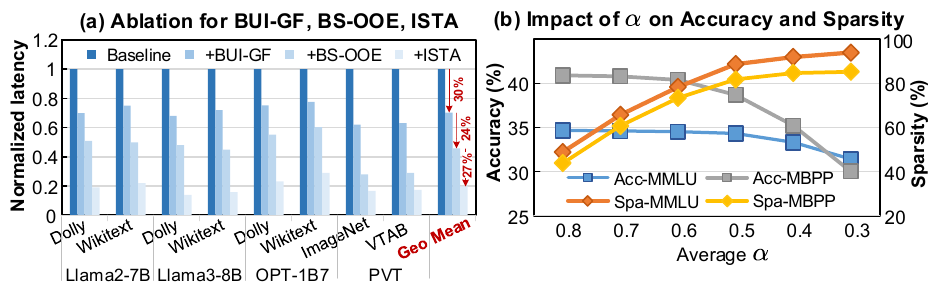}\vspace{-3mm}
\caption{(a) Latency reduction for BUI-GF, BS-OOE and ISTA. (b) Exploring the trade-off between accuracy and sparsity.}
\label{fig:latency_break_down}\vspace{-1mm}
\end{figure}

\subsection{Algorithm Performance}\label{subsec:Algorithm_Performance} 
\textbf{Algorithm settings}: INT8 models serve as the accuracy baseline, and $\alpha$ (Eq.\eqref{eq:Threshold}) is adjusted in 0.1 increments to evaluate the accuracy and overhead for each benchmark. Two PADE configurations are evaluated: standard ($0\%$ loss), aggressive ($1\%$ loss).

Fig. \mbox{\ref{fig:SW-SparseAttention}} (a)(b) compares the accuracy of three representative software-only sparse attention methods, two predictor-free works (SpAtten \cite{wang2021spatten}, DTATrans \cite{yang2022dtatrans}), and PADE. The \textit{Sparsity Level} denotes the ratio between the computation cost of sparse execution (prediction + actual computation) and that of dense execution. StreamingLLM \cite{xiao2023efficient} adopts a static sparsity pattern—retaining only the initial and recent tokens, while MInference \cite{jiang2024minference}, and DoubleSparsity \cite{yang2024post} rely on runtime sparsity prediction. Four key observations are made: \textbf{1)} StreamingLLM performs the worst, due to its lack of adaptivity in capturing input-dependent sparsity patterns. \textbf{2)} MInference improves accuracy by combining dynamic prediction with predefined sparsity patterns but remains limited by restricted pattern diversity. \textbf{3)} DoubleSparsity introduces a more flexible dynamic sparsity mechanism and integrates channel sparsity to reduce prediction overhead. However, as its prediction computation and memory access cannot be reused in subsequent steps, it suffers from inefficiency, yielding slightly lower accuracy than PADE at the same sparsity level. \textbf{4)} For SpAtten and DTATrans, comparable accuracy to PADE can be achieved only after fine-tuning. This is mainly because both methods eliminate the predictor by using the previous layer’s attention distribution to guide pruning in the current layer, which introduces significant errors without fine-tuning. In contrast, PADE consistently achieves the best performance across all settings, benefiting from its fine-grained sparsity prediction and efficient reuse of both computation and memory access.

Fig. \ref{fig:SW-SparseAttention}(c) further compares PADE (software–hardware co-optimization) with the software-only methods under the same 1\% accuracy loss. PADE achieves an average 5.2$\times$ speedup and 10.4$\times$ improvement in energy efficiency. Moreover, PADE’s advantage becomes more pronounced as the sequence length increases, since longer sequences amplify the overhead of sparsity prediction. In such case, PADE’s stage-fusion and reuse mechanisms effectively eliminate this prediction overhead, leading to superior scalability and efficiency.

\begin{figure}[t]
\centering
\includegraphics[width=\linewidth]{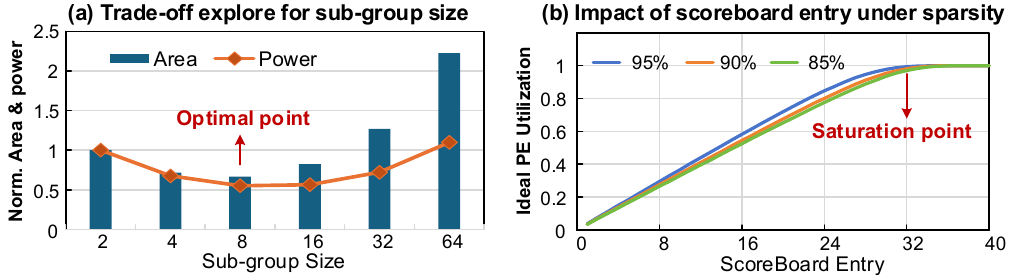}\vspace{-3mm}
\caption{Exploration of optimal (a) Sub-group size. (b) Scoreboard entry. }
\label{fig:hardware-trade-off}\vspace{-2mm}
\end{figure}

\textbf{Computation Reduction}. Fig. \ref{fig:Overall_Complexity} compares the computation reduction across accelerators with $0\%$ accuracy loss. Spatten, without retraining, yields the lowest reduction and serves as the baseline. Energon enhances performance via progressive precision prediction, achieving a $32\%$ reduction, and outperforming coarse-grained prediction accelerators like Sanger and DOTA. However, it lacks computation reuse. SOFA mitigates this by using log-domain differential leading-one computation, but still relies on an added predictor, limiting its reduction to $45\%$. In contrast, PADE eliminates the need for a predictor by leveraging fine-grained bit-level early termination and bit reuse, achieving a reduction of up to $71.6\%$.

\textbf{Memory Access Reduction}. As shown in Fig. \ref{fig:Overall_Complexity}, Sanger, which employs an extra 4-bit MSB predictor, serves as the baseline. DOTA adopts low-rank approximation but fails to mitigate prediction bitwidth overhead. Energon and Spatten* (with finetune) partially alleviate this via mixed precision, but still rely on an extra predictor, limiting their reduction to $21\%$ and $42\%$, respectively. SOFA reduces memory overhead via cross-stage tiling yet remains constrained by the extra predictor. In contrast, PADE achieves an average memory reduction of $75.8\%$ across both long- and short-sequence tasks, attributed to the BSF strategy. When compared to Spatten without fine-tune, PADE achieves $3.4\times$ higher memory reduction.

\subsection{Design of Architecture Parameters}
The key architectural parameters in PADE are decided by detailed workload profiling and DSE exploration.

\textbf{(1) Computational throughput}: The INT-8 throughput ratio between QK-PU and V-PU is set to 8:1 (see Table \ref{tab:PADE_Architecture}), derived from the typical QK-to-SV computation ratio observed in LLM workloads due to sparsity.

\textbf{(2) Buffer Sizes}: A 320 KB KV buffer and a 32 KB query buffer are allocated, sufficient for 12.8k tokens under typical sparsity ratio of 0.2 (64-d embedding) and 256 queries, respectively. These configurations balance on-chip storage efficiency and multi-phase processing for longer sequences.  


\textbf{(3) Optimal Sub-group Size}. We conduct a DSE to determine the optimal PE group size (Recall \mbox{\S\ref{subsec:ANDer_Tree}}). As shown in Fig. \mbox{\ref{fig:hardware-trade-off}} (a), a sub-group size of 8 minimizes area and power overhead, which is therefore adopted in our PADE accelerator.

\textbf{(4) Optimal Scoreboard Size}. A larger scoreboard can cache more partial sums, which improves PE utilization but increases area overhead. Fig. \ref{fig:hardware-trade-off} (b) shows that PE utilization saturates at around 32 entries. This indicates that, at this point, memory access and computation rates are balanced, preventing PE computation from being blocked. Therefore, PADE adopts a 32-entry scoreboard, as in Table \ref{tab:PADE_Architecture}.



\subsection{Architecture Evaluation}\label{subsec:Architecture_Evaluation}
\textbf{Ablation}. We conduct an ablation study to evaluate the latency reduction of BUI-GF, BS-OOE, and ISTA against a baseline dense attention accelerator derived from PADE, but without sparse processing modules. As shown in Fig. \ref{fig:latency_break_down} (a), BUI-GF reduces average latency by $30\%$, mainly by predictor-free token sparsity. Further, BS-OOE realizes $24\%$ latency reduction via improved hardware utilization. Finally, ISTA achieves $27\%$ decrease, via efficient tiling and data reuse.

\begin{figure}[t]
\includegraphics[width=1.0\linewidth]{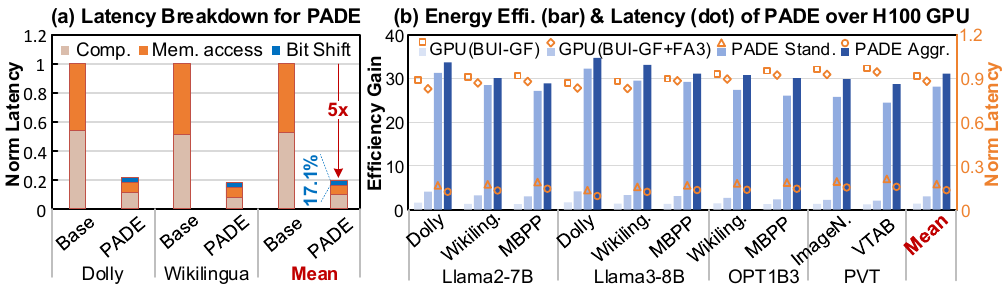}\vspace{-3mm}
\caption{(a) Breakdown for the overhead of bit shifting. (b) Latency and Efficiency gain of PADE over Nvidia H100 GPU.} 
\label{fig:GPU_Throughput_Gain}\vspace{-2mm}
\end{figure}

\textbf{Accuracy $\&$ Sparsity Trade-off}. 
The BUI-GF may affect accuracy, as it introduces a parameter $\alpha$ to pruning KVs (\S\ref{subsec:BUI-GF}). Fig. \ref{fig:latency_break_down} (b) shows the impact of $\alpha$ on accuracy and sparsity using LLaMA2-7B on MMLU (reasoning) and MBPP (generation). Overall, a smaller $\alpha$ results in more aggressive pruning, decreasing accuracy but increasing sparsity. There are some key observations: For generation tasks (MBPP), accuracy drops noticeably when $\alpha$$<$$0.6$. In contrast, for reasoning tasks (MMLU), the model is more tolerant to pruning, with accuracy degrading evidently only when $\alpha$$<$$0.5$. This may be because reasoning tasks rely on vital tokens for inference, resulting in higher token redundancy. On the other hand, the sparsity gains begin to diminish when $\alpha$$<$$0.5$, likely due to overly aggressive pruning, which harms crucial tokens and limits further sparsity. Therefore, to strike a balanced trade-off between accuracy and sparsity, we empirically set $\alpha$ within the range of 0.5–0.6.


\textbf{Bit-serial overhead}. Fig.\ref{fig:GPU_Throughput_Gain} (a) profiles latency overhead between the PADE architecture with value-level INT8 computation (baseline) and the bit-level PADE design. The 17\% bit-shifting overhead is outweighed by a 5$\times$ latency reduction, validating the effectiveness of bit-level optimizations.

\textbf{Comparison with GPU}.Fig. \mbox{\ref{fig:GPU_Throughput_Gain}} (b) compares the latency and energy efficiency of PADE with H100 GPU across various benchmarks. As shown, even with sparsity detection, BUI-GF enables only an average of $8\%$ latency reduction and 1.3$\times$ efficiency gain on the GPU. Incorporating FlashAttention3 improves these figures to 14\% latency reduction and 3.1$\times$ efficiency gain through memory-access reduction via tiling, but the improvement remains limited. This is because GPUs cannot leverage fine-grained bit-grained early termination or bit sparsity, nor can they efficiently execute bit-wise out-of-order computation. By contrast, PADE achieves an average $78\%$ utilization thanks to dedicated hardware support, such as scoreboard PE, BS ANDer tree, and an efficient pipeline between QK-PU and V-PU. Overall, PADE standard and aggressive achieve an average 5.8$\times$/7.4$\times$ latency reductions and 28.2$\times$/31.1$\times$ energy efficiency gains, respectively.



\begin{figure}[t]
\centering
\includegraphics[width=\linewidth]{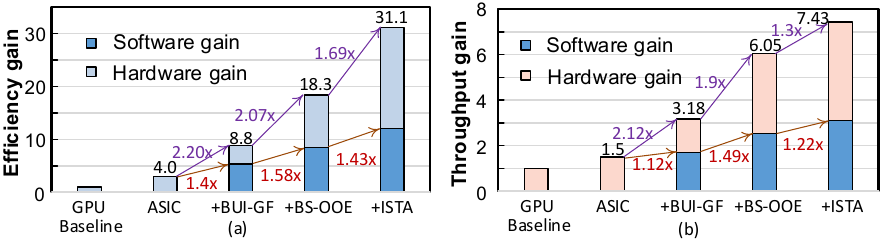}\vspace{-3mm}
\caption{Energy efficiency and throughput gain breakdown.}
\label{fig:Break_down}\vspace{-5mm}
\end{figure}

\textbf{Efficiency gain breakdown.} Fig. \ref{fig:Break_down} (a) gives the energy efficiency breakdown. With a dedicated ASIC and customized datapath, PADE achieves a $4.0\times$ gain over the GPU baseline. While the token sparsity-leveraged BUI-GF theoretically reduces computation by $3.7\times$, practical efficiency improves by only $1.4\times$. This is due to redundantly loading repeated bit planes. After adding the dedicated scoreboard-based result reusable PE lane, the performance jumps by $2.2\times$. Similarly, directly applying the BS-OOE scheme and tiling-target ISTA scheme yields only $1.58\times$ and $1.43\times$ efficiency gain. This is due to mismatched computational granularity and the presence of unused V vectors, which ultimately lead to severe resource underutilization. By contrast, deploying tailored engines can further bring $2.07\times$ and $1.69\times$ efficiency gain effects.

\begin{figure}[t]
\centering
\includegraphics[width=0.96\linewidth]{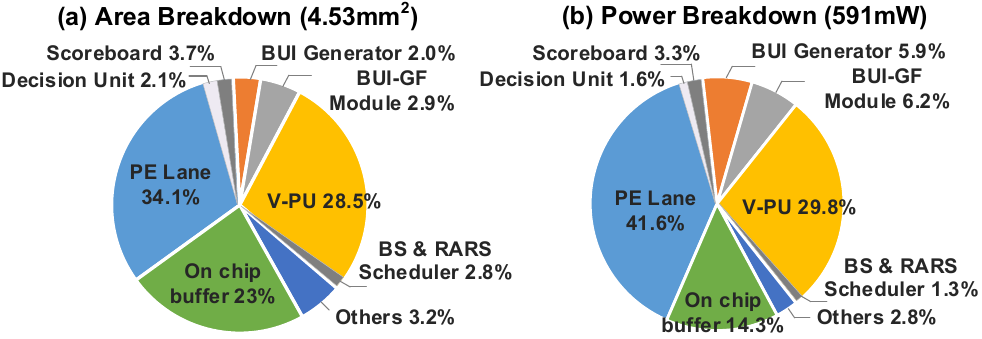}\vspace{-2mm}
\caption{Area/Power of PADE at TSMC 28nm, 800MHz.}
\label{fig:Area_Power}\vspace{-2mm}
\end{figure}

\textbf{Area and Power}. Fig. \ref{fig:Area_Power} presents the area and power breakdown of PADE. Occupying 4.53 mm$^2$ and consuming 591 mW, PADE achieves a peak energy efficiency of 11.36 TOPS/W. The added BUI Generator and BUI-GF modules adaptively respond to attention distribution for token pruning, incurring only $4.9\%$ area and $12.1\%$ power overhead. Also, integrating the Scoreboard and Decision Unit into PE lanes enables stage fusion with just $5.8\%$ area and $4.9\%$ power cost. Despite the modest overhead, eliminating the sparsity predictor and reducing off-chip memory access yield notable speedup and efficiency gains. As depicted in Fig.\ref{fig:Break_down} (b), compared to baseline without sparse processing modules, software-hardware co-design BUI-GF, BS-OOE and ISTA bring $2.12\times$, $1.9\times$ and $1.3\times$ throughput gain, respectively. In summary, PADE represents a deliberate tradeoff, achieving substantial efficiency gains with minimal resource overhead.

\textbf{Data Layout.} In PADE, the data layout is carefully co-designed across the off-chip DRAM and on-chip SRAM to maximize memory bandwidth utilization. As depicted in Fig. \mbox{\ref{fig:Memory_Layout}} above, \textbf{\textit{K is bank-interleaved along the bit dimension}}, meaning that each DRAM bank stores a distinct bit-plane of the K tensor, enabling efficient bit-plane access when performing bit-level computations. In contrast,\textbf{\textit{ Q and V are bank-interleaved along the hidden (H) dimension}} such that 8-bit data is read continuously. When the data are fetched into the on-chip SRAM, the layout is reorganized to match the PE access pattern. In the Q,V SRAM, each row primarily stores different bits of the same element, whereas in the K SRAM, each row stores the same bit plane (e.g., MSB) from multiple elements across the hidden dimension.

\subsection{Comparison to SOTA Accelerators}\label{subsec:Comparison_to_SOTA_Accelerators}
\begin{figure}[t]
\centering
\includegraphics[width=\linewidth]{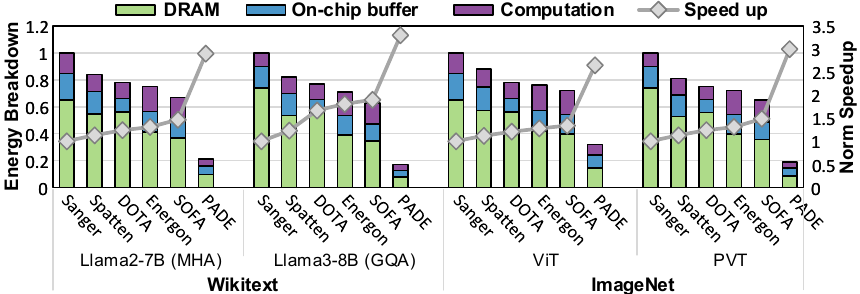}\vspace{-2mm}
\caption{Speedup and energy breakdown comparisons.}
\label{fig:Speedup_and_energy_breakdown}\vspace{-2mm}
\end{figure}

Fig. \ref{fig:Speedup_and_energy_breakdown} compares the throughput and energy efficiency of PADE with five SOTA attention accelerators. Energy overhead is decomposed into computation, on-chip buffer, and off-chip memory. All existing designs rely on additional predictors to estimate attention sparsity. While these works partially reduce computational overhead, the inclusion of additional sparsity predictors limits their efficiency gains. Moreover, most designs fail to reduce off-chip memory access costs, leading to DRAM consistently accounting for over $65\%$ of total energy. Two key observations are made: \textbf{1)} Llama2-7B vs. Llama3-8B: PADE achieves greater performance gains when GQA is adopted, as the scoreboard-based PE enhances key reuse across heads. \textbf{2)} ViT vs. PVT: PADE’s acceleration advantage grows with longer sequences, as higher sparsity amplifies the predictor overhead in conventional designs, causing performance degradation. In contrast, PADE's predictor-free architecture avoids this overhead entirely. Overall, PADE achieves the highest performance across all workloads, achieving average speedups of 3$\times$, 2.2$\times$, $1.9\times$ over Sanger, DOTA and SOFA, respectively, along with energy savings of 5.1$\times$, 4.3$\times$, 3.4$\times$.    



\begin{figure}[t]
\centering
\includegraphics[width=0.95\linewidth]{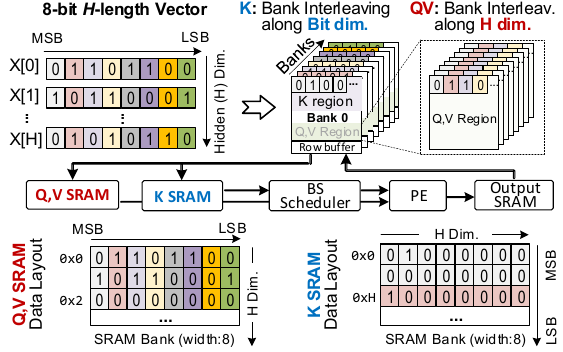}\vspace{-3mm}
\caption{DRAM, SRAM layouts and overall dataflow of PADE.}
\label{fig:Memory_Layout}\vspace{-2mm}
\end{figure}

\begin{figure}[t]
\centering
\includegraphics[width=\linewidth]{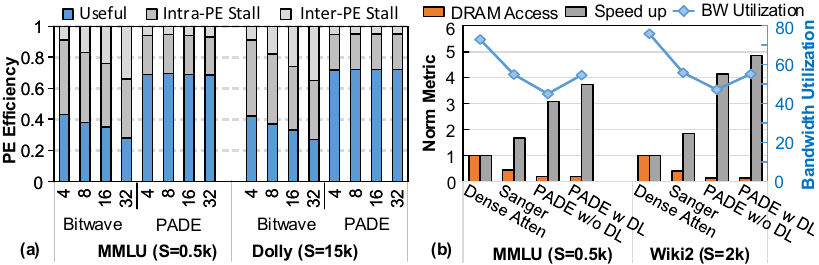}\vspace{-3mm}
\caption{(a) Breakdown of execution cycles versus the number of PE lanes. (b) Bandwidth utilization, DRAM access, and speedup comparisons.}
\label{fig:Workload_Balance}\vspace{-2mm}
\end{figure}

\textbf{Workload Balance}. Fig. \mbox{\ref{fig:Workload_Balance}} (a) presents a detailed breakdown of execution time across varying numbers of PE lanes to illustrate load balance effects. To better show PADE's advantages, we compare it against a SOTA bit accelerator, BitWave \mbox{\cite{shi2024bitwave}}, which leverages bit-flipping to enhance bit-plane sparsity. Since each PE lane integrates multiple bit-serial multipliers, intra-PE stall arises when certain multipliers handle more effective bits. Inter-PE install, in contrast, results from variations in bit sparsity across different key vectors. As the number of PE lanes increase, BitWave suffers from greater intra- and inter-PE imbalance, as it only exploits bit-0 sparsity, which exhibits large variability. In comparison, PADE adopts a more balanced bit sparsity distribution, thereby achieving around $30\%$ higher PE utilization.

\textbf{DRAM Bandwidth (BW) Utilization.} Fig. \mbox{\ref{fig:Workload_Balance}} (b) compares DRAM access, speedup, and bandwidth (BW) utilization of different attention accelerators on the MMLU and Wikitext2 workloads. In this context, \textit{Dense Attention} refers to the method without sparse computation, while \textit{Sanger} employs a coarse-grained 4-bit value for sparse prediction. \textit{PADE w/o DL} represents PADE's fine-grained bit prediction without data layout optimization in DRAM and SRAM. \textit{PADE w DL} refers to PADE with customized bit-oriented data layout, as shown in Fig. \mbox{\ref{fig:Memory_Layout}}. Compared to the dense version, although PADE's bit-grained sparsity lowers DRAM bandwidth utilization by around 30\%, memory access decreases over 6.7$\times$, resulting in an average 3.4$\times$ speedup. After incorporating the bit-oriented data layout, the BW utilization improves to 58\% due to higher row buffer hits, achieving a speedup of 4.3$\times$.

\subsection{Discussion for Deployment and Scalability}\label{subsec:discussion}
\begin{figure}[t]
\centering
\includegraphics[width=\linewidth]{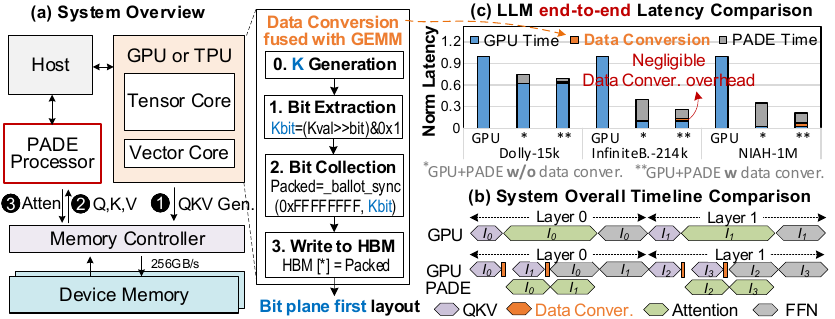}\vspace{-2mm}
\caption{Conceptual integration of PADE with a GPU-like NN accelerator.}
\label{fig:System_Integration}\vspace{-4mm}
\end{figure}

\textbf{System Integration}. PADE functions as a co-processor working collaboratively with the GPU. The GPU handles dense computations such as QKV projection and FFN, while PADE accelerates attention via sparsity. As depicted in Fig. \ref{fig:System_Integration} (a), both processors execute instructions issued by the host CPU and share the device memory, enabling direct data exchange without additional transfers. During the K generation, GPU executes an extra data conversion operation to store K in HBM with a bit-plane-first layout, as shown in Fig. \ref{fig:Memory_Layout}.

Fig. \ref{fig:System_Integration} (b) compares the overall execution timeline for the PADE-equipped GPU (P-G) system with the original GPU-only system. As shown, the operations concerning two successive input sequences ($I_0$ and $I_1$) are interleaved on both GPU and PADE processors, greatly improving the system throughput. Fig. \ref{fig:System_Integration} (c) quantifies this speedup, showing that at 214k scenarios, the P-G system achieves a 2.1$\times$ speedup. However, without a customized data layout, redundant memory accesses limit performance gains. After incorporating the bit-oriented data layout, despite a latency increase of less than 2\%, an additional 1.9$\times$ speedup is achieved.

\begin{figure}[t]
\centering
\includegraphics[width=\linewidth]{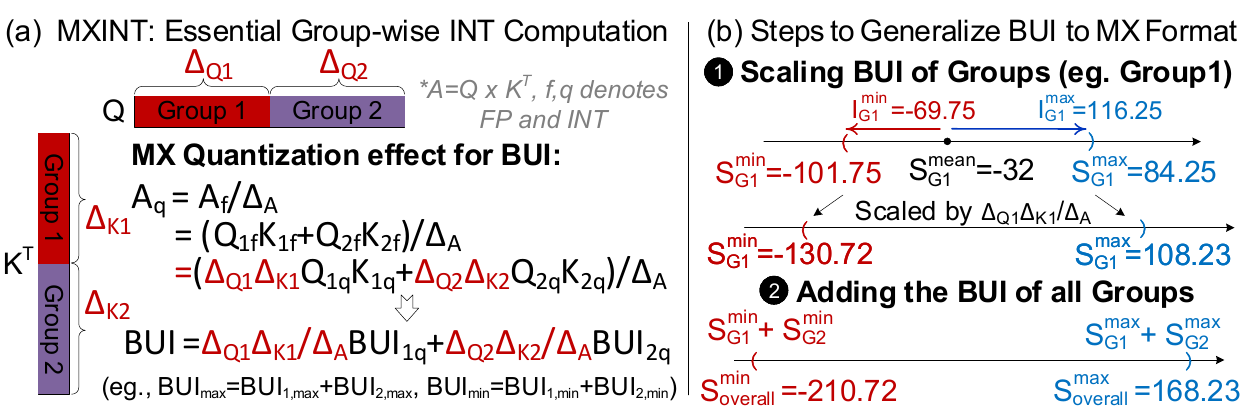}\vspace{-2mm}
\caption{Compatibility of the BUI-GF mechanism with the MX format.}
\label{fig:miro-scale-format}\vspace{-2mm}
\end{figure}

\textbf{Extension for MXINT.} The micro-scale format performs fine-grained quantization along the channel dimension by grouping data into 32-element segments \cite{rouhani2023microscaling}. PADE ensures compatibility by applying group-wise scaling to the bit uncertain interval (BUI) (\S\ref{subsec:BUI-GF}). As shown in Fig. \ref{fig:miro-scale-format} (a), when processing 64-length Q and K vectors, the micro-format partitions them into two 32-element groups and quantizes each group using calibration-derived factors ($\Delta$), resulting in group-wise BUI scaling. As depicted in Fig. \ref{fig:miro-scale-format} (b), PADE achieves compatibility via two steps. \ding{182} First, it performs bitwise serial multiplication within each 32-length group, with the BUI derived from the strategy in 
\S\ref{subsec:BUI-GF}. The resulting BUI is then expanded using the quantization factor, eg., $\Delta_{\rm Q1}\Delta_{\rm K1}/\Delta_{\rm A}$. \mbox{\ding{183}} Finally, the max and min BUI values across all groups are aggregated to compute the overall BUI. This method can be extended to dot products of arbitrary length. 

\textbf{Extension for FP formats.} Prior studies \cite{sheng2023flexgen,xiao2023smoothquant,zhao2024atom} have demonstrated that the K and V tensors are highly amenable to quantization because the softmax normalization in self-attention naturally suppresses quantization noise. This property enables safe quantization to INT8 or even INT4 with negligible accuracy degradation. Motivated by this, when queries operate in FP format, PADE converts the INT-FP computation into a bit-serial form through exponent alignment, following methodologies adopted in prior works \cite{chen2025bitmod,jang2024figna,fang2025anda}. The resulting bit-serial execution is fully compatible with PADE’s existing processing mechanism.

\begin{figure}[t]
\centering
\includegraphics[width=\linewidth]{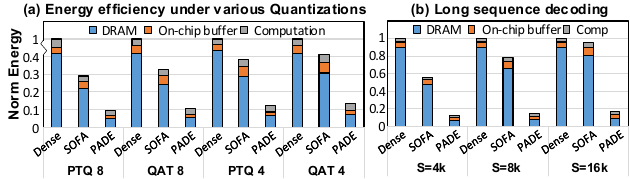}\vspace{-3mm}
\caption{Efficiency with (a) diverse quantizations, (b) long sequence decoding.}
\label{fig:Scalability}\vspace{-4mm}
\end{figure}

\textbf{Diverse Quantizations}. To assess PADE's adaptability across quantization strategies and bit-widths, we extend both PADE and SOFA to QAT/PTQ INT8 and INT4 scenarios, as shown in Fig.\ref{fig:Scalability} (a). There are two key observations: (1) Compared to PTQ8, QAT8 leads to an ~$6\%$ increase in energy consumption for SOFA. This is due to the more uniform data distribution under QAT, which reduces sparsity and renders SOFA’s predictor largely ineffective. In contrast, PADE also shows a slight energy increase, but the overhead remains negligible, as it avoids the use of any explicit predictor. (2) When the bit-width is reduced to 4 bits, SOFA's energy efficiency gain drops significantly, as the predictor becomes the dominant source of power consumption under low-precision settings. In contrast, PADE’s energy gain decreases by only $2\%$, thanks to its predictor-free design that avoids this overhead. 

\textbf{Ultra-long Sequence Decoding}. In long-sequence decoding, memory access optimization is critical due to the lack of data reuse, making it a key indicator of a design's memory efficiency. As shown in Fig. \ref{fig:Scalability} (b), in decoding, DRAM access accounts for over $85\%$ of total power overhead across all designs, due to the autoregressive nature of the workload. Notably, compared to the dense version, SOFA's energy consumption increases significantly with sequence length—rising by nearly $40\%$ from 4K to 16K tokens. This is because SOFA must load all key vectors corresponding to the predicted sequence at each decoding step, making the predictor's overhead grow rapidly. In contrast, PADE shows only a modest increase of about $5\%$ over the same sequence range, thanks to its predictor-free architecture. \textbf{These results collectively demonstrate the advantages of PADE across diverse scenarios.}

\subsection{Limitations and Future Direction}
While PADE significantly advances sparse-attention acceleration by eliminating the sparsity predictor through unified bit-serial stage fusion, several limitations remain for future works: (1) As the sequence length of large language models (LLMs) continues to scale, the need for distributed attention becomes inevitable. Extending PADE to distributed scenarios, especially in emerging wafer-scale architectures \cite{wang2026WATOS,wang2026TEMP,wang2024tmac,he2025waferllm,hu2024wafer,tang2025moentwine}, remains an open problem. (2) Beyond single-bit granularity, exploring multi-bit stage fusion may provide a more favorable efficiency–accuracy trade-off, which represents a promising direction for future PADE enhancements.

\section{Related Works}
\textbf{Efficient Attention Accelerators}. Numerous attention accelerators \cite{wang2025bitstopper,wang2025lapa,wang2025beta,li2020ftrans, ham20203, ham2021elsa, qu2022dota, fan2023taskfusion, yang2022dtatrans, hong2022dfx, liu2024hsconn, wang2024sofa, wang2022energy, fang2022algorithm, yazdanbakhsh2022sparse, li2022accelerating, lu2021sanger, wang2021spatten, zhou2022energon, qin2023fact, shen2022salo, fan2022adaptable, bai2024swat,zhao2024hardware,tuli2023acceltran,you2023vitcod,shen2022salo} have been proposed. Early works, such as A$^3$ \cite{ham20203} and ELSA \cite{ham2021elsa} accelerate computation via approximation techniques. Recent efforts have shifted to jointly optimizing computation and memory. Energon \cite{zhou2022energon} and SOFA \cite{wang2024sofa} adopt fine-grained filtering and FlashAttention-optimized tiling techniques to alleviate memory overhead. However, they still rely on extra sparsity predictors, which become de facto latency and power bottlenecks after sparsification. PADE is the first to explicitly identify and address this issue. It removes the need for external predictors by fine-grained, bit-serial computation, while leveraging efficient bit-level early termination and reuse. 




\textbf{Neural network (NN) accelerator with sparsity}. Numerous accelerators~\cite{wang2021efficient,hojabr2021spaghetti,wang2018low,gondimalla2019sparten,asgari2020alrescha,deng2021gospa,qin2020sigma,gudaparthi2022candles,fang2022algorithm,hanson2022cascading,lew2022anticipating,li2022ristretto,li2021escalate,li2022accelerating,liu2022s2ta,pavon2021via,rucker2021capstan,sadi2019efficient,walia2021fast,yang2020procrustes,Gon2023Eureka,wu2022sparseloop,wu2023highlight,shin2022griffin} exploit sparsity to accelerate NN inference. General-purpose sparse tensor accelerators \cite{srivastava2020matraptor,mahmoud2020tensordash,kwon2019tensordimm,hegde2019extensor,kanellopoulos2019smash,chen2020tpspmv} support operations on sparse fully connected layers, but most of them target pre-trained, statically sparse weights. By contrast, PADE targets attention dynamic sparsity, which requires on-the-fly prediction, making zero-based sparsity methods ineffective. While some recent works explore activation sparsity \cite{jang2021sparsity} or combine weight and activation sparsity \cite{wu2023highlight,huang2023rm,wang2021dual}, they still rely on zero-based sparsity, failing to address the argmax sparsity, which is the PADE targets.

\textbf{Bit-serial computing accelerators}.
Prior works \cite{wang2025mcbp,im2023sibia,han2024bitnn,yang2021fusekna,judd2016stripes,albericio2017bit,delmas2019bit,lo2023bit,sharify2019laconic,lee2018unpu,delmas2017dynamic,gondimalla2019sparten,li2022ristretto,parashar2017scnn,yang2023isosceles,lu2021distilling,kam2025panacea,li2025lut} accelerates NNs by exploiting bit-level sparsity \cite{albericio2017bit,delmas2019bit,lu2021distilling,sharify2019laconic,yang2021fusekna} or dynamically reducing bit-width \cite{gondimalla2019sparten,li2022ristretto,parashar2017scnn,yang2023isosceles}. However, these techniques are difficult to apply directly to sparse attention, as they typically rely on offline weight preprocessing, while sparse attention demands real-time, token-wise prediction. Further, the inherently low utilization of bit-level prediction further limits their effectiveness in attention acceleration. In contrast, PADE adopts a lightweight runtime pruning strategy, BUI-GF, and maximizes hardware efficiency via bidirectional bit sparsity and streamlined out-of-order execution.

\section{Conclusion}
We propose PADE, a software-hardware co-design for dynamic sparse attention without relying on traditional sparsity predictors. Through bit-level early termination, dual-sided bit sparsity, dedicated bit-wise out-of-order execution, and an optimized tiling dataflow, PADE achieves up to $31.1\times$ energy efficiency over H100 and $5.1\times$ over SOTA accelerator Sanger.

\section*{Acknowledgments}
This work was supported in part by the National Science and Technology Major Project under Grant 2022ZD0115200; in part by the NSFC under Grant 62125403, Grant U24A20234, Grant 92464302 and Grant U24B20164; in part by the Beijing S\&T Project Z251100008425010; in part by Shanghai Municipal Science and Technology Major Project; the Natural Science Foundation of Jiangsu Province Basic Research Program under Grant BK20243042; in part by the Beijing National Research Center for Information Science and Technology; in part by the Northern IC Technology Innovation Center (Beijing) Co., Ltd under Grant QYJS20232801B; and in part by the Beijing Advanced Innovation Center for Integrated Circuits.


\bibliographystyle{IEEEtranS}
\bibliography{main}

@String{BIT = "{BIT}" }

@String{Computing = "Computing" }

@String{Computer = "{IEEE} Computer" }

@article{shah2024flashattention,
  title={Flashattention-3: Fast and accurate attention with asynchrony and low-precision},
  author={Shah, Jay and Bikshandi, Ganesh and Zhang, Ying and Thakkar, Vijay and Ramani, Pradeep and Dao, Tri},
  journal={Advances in Neural Information Processing Systems},
  volume={37},
  pages={68658--68685},
  year={2024}
}

@inproceedings{song2018situ,
  title={{In-situ AI: Towards autonomous and incremental deep learning for IoT systems}},
  author={Song, Mingcong and Zhong, Kan and Zhang, Jiaqi and Hu, Yang and Liu, Duo and Zhang, Weigong and Wang, Jing and Li, Tao},
  booktitle={2018 IEEE International Symposium on High Performance Computer Architecture (HPCA)},
  pages={92--103},
  year={2018},
  organization={IEEE}
}

@ArtifactSoftware{R,
    title = {R: A Language and Environment for Statistical Computing},
    author = {{R Core Team}},
    organization = {R Foundation for Statistical Computing},
    address = {Vienna, Austria},
    year = {2019},
    url = {https://www.R-project.org/},
}

@inproceedings{snyder2004verilator,
  title={{Verilator and SystemPerl}},
  author={Snyder, Wilson},
  booktitle={North American SystemC Users’ Group, Design Automation Conference},
  year={2004}
}

@inproceedings{shi2024bitwave,
  title={BitWave: Exploiting Column-Based Bit-Level Sparsity for Deep Learning Acceleration},
  author={Shi, Man and Jain, Vikram and Joseph, Antony and Meijer, Maurice and Verhelst, Marian},
  booktitle={Proceedings of IEEE International Symposium on High-Performance Computer Architecture (HPCA)},
  pages={732--746},
  year={2024}
}

@article{brown2020language,
  title={{Language models are few-shot learners}},
  author={B. Brown, Tom and Mann, Benjamin and Ryder, Nick and Subbiah, Melanie and Kaplan, Jared D and Dhariwal, Prafulla and Neelakantan, Arvind and Shyam, Pranav and Sastry, Girish and Askell, Amanda and  Agarwal, Sandhini and Herbert-Voss, Ariel and Krueger, Gretchen and Henighan, Tom and Child, Rewon and Ramesh, Aditya and M. Ziegler, Daniel and Wu, Jeffrey and Winter, Clemens and Hesse, Christopher and Chen, Mark and  Sigler, Eric and Litwin, Mateusz and Gray, Scott and Chess, Benjamin and Clark, Jack and Berner, Christopher and McCandlish, Sam and Radford, Alec and Sutskever, Ilya and Amodei, Dario},
  journal={Advances in neural information processing systems},
  volume={33},
  pages={1877--1901},
  year={2020}
}

@inproceedings{geng2019o3bnn,
  title={{O3BNN: An out-of-order architecture for high-performance binarized neural network inference with fine-grained pruning}},
  author={Geng, Tong and Wang, Tianqi and Wu, Chunshu and Yang, Chen and Wu, Wei and Li, Ang and Herbordt, Martin C},
  booktitle={Proceedings of the ACM International Conference on Supercomputing},
  pages={461--472},
  year={2019}
}

@article{fuad2023survey,
  title={{A survey on sparsity exploration in Transformer-based accelerators}},
  author={Fuad, Kazi Ahmed Asif and Chen, Lizhong},
  journal={Electronics},
  volume={12},
  number={10},
  pages={2299},
  year={2023},
  publisher={MDPI}
}

@inproceedings{o2017fine,
  title={{Fine-grained DRAM: Energy-efficient DRAM for extreme bandwidth systems}},
  author={O'Connor, Mike and Chatterjee, Niladrish and Lee, Donghyuk and Wilson, John and Agrawal, Aditya and Keckler, Stephen W and Dally, William J},
  booktitle={Proceedings of the 50th Annual IEEE/ACM International Symposium on Microarchitecture},
  pages={41--54},
  year={2017}
}

@article{zhao2024atom,
  title={{Atom: Low-bit quantization for efficient and accurate LLM serving}},
  author={Zhao, Yilong and Lin, Chien-Yu and Zhu, Kan and Ye, Zihao and Chen, Lequn and Zheng, Size and Ceze, Luis and Krishnamurthy, Arvind and Chen, Tianqi and Kasikci, Baris},
  journal={Proceedings of Machine Learning and Systems},
  volume={6},
  pages={196--209},
  year={2024}
}

@article{yang2024post,
  title={Post-training sparse attention with double sparsity},
  author={Yang, Shuo and Sheng, Ying and Gonzalez, Joseph E and Stoica, Ion and Zheng, Lianmin},
  journal={arXiv preprint arXiv:2408.07092},
  year={2024}
}

@article{wang2025beta,
  title={{BETA: A bit-Grained Transformer attention accelerator With efficient early termination}},
  author={Wang, Huizheng and Wang, Hongbin and Yue, Zhiheng and Liu, Jingyao and Wei, Taiquan and Hu, Shaojun Wei Yang and Yin, Shouyi},
  journal={IEEE Transactions on Circuits and Systems II: Express Briefs},
  year={2025},
  publisher={IEEE}
}

@article{frantar2022gptq,
  title={{GPTQ: Accurate post-training quantization for generative pre-trained transformers}},
  author={Frantar, Elias and Ashkboos, Saleh and Hoefler, Torsten and Alistarh, Dan},
  journal={arXiv preprint arXiv:2210.17323},
  year={2022}
}

@article{li2024large,
  title={Large language model inference acceleration: A comprehensive hardware perspective},
  author={Li, Jinhao and Xu, Jiaming and Huang, Shan and Chen, Yonghua and Li, Wen and Liu, Jun and Lian, Yaoxiu and Pan, Jiayi and Ding, Li and Zhou, Hao},
  journal={arXiv preprint arXiv:2410.04466},
  year={2024}
}

@article{qiu2019blockwise,
  title={Blockwise self-attention for long document understanding},
  author={Qiu, Jiezhong and Ma, Hao and Levy, Omer and Yih, Scott Wen-tau and Wang, Sinong and Tang, Jie},
  journal={arXiv preprint arXiv:1911.02972},
  year={2019}
}

@inproceedings{dong2022cswin,
  title={Cswin Transformer: A general vision Transformer backbone with cross-shaped windows},
  author={Dong, Xiaoyi and Bao, Jianmin and Chen, Dongdong and Zhang, Weiming and Yu, Nenghai and Yuan, Lu and Chen, Dong and Guo, Baining},
  booktitle={Proceedings of the IEEE/CVF conference on computer vision and pattern recognition},
  pages={12124--12134},
  year={2022}
}

@article{child2019generating,
  title={Generating long sequences with sparse transformers},
  author={Child, Rewon and Gray, Scott and Radford, Alec and Sutskever, Ilya},
  journal={arXiv preprint arXiv:1904.10509},
  year={2019}
}

@inproceedings{parmar2018image,
  title={Image transformer},
  author={Parmar, Niki and Vaswani, Ashish and Uszkoreit, Jakob and Kaiser, Lukasz and Shazeer, Noam and Ku, Alexander and Tran, Dustin},
  booktitle={International conference on machine learning},
  pages={4055--4064},
  year={2018},
  organization={PMLR}
}

@article{dettmers2022gpt3,
  title={Gpt3. int8 (): 8-bit matrix multiplication for transformers at scale},
  author={Dettmers, Tim and Lewis, Mike and Belkada, Younes and Zettlemoyer, Luke},
  journal={Advances in Neural Information Processing Systems},
  volume={35},
  pages={30318--30332},
  year={2022}
}

@article{cavigelli2016origami,
  title={{Origami: A 803-GOp/s/W convolutional network accelerator}},
  author={Cavigelli, Lukas and Benini, Luca},
  journal={IEEE Transactions on Circuits and Systems for Video Technology},
  volume={27},
  number={11},
  pages={2461--2475},
  year={2016},
  publisher={IEEE}
}

@article{bai2024seed,
  title={{Seed-ASR: Understanding diverse speech and contexts with LLM-based speech recognition}},
  author={Bai, Ye and Chen, Jingping and Chen, Jitong and Chen, Wei and Chen, Zhuo and Ding, Chuang and Dong, Linhao and Dong, Qianqian and Du, Yujiao and Gao, Kepan},
  journal={arXiv preprint arXiv:2407.04675},
  year={2024}
}

@inproceedings{ladhak-wiki-2020,
    title={{WikiLingua: A new benchmark dataset for multilingual abstractive summarization}},
    author={Faisal Ladhak, Esin Durmus, Claire Cardie and Kathleen McKeown},
    booktitle={Findings of EMNLP, 2020},
    year={2020}
}

@article{le2022bloom,
  title={{Bloom: A 176B-parameter open-access multilingual language model}},
  author={Le Scao, Teven and Fan, Angela and Akiki, Christopher and Pavlick, Ellie and Ili{\'c}, Suzana and Hesslow, Daniel and Castagn{\'e}, Roman and Luccioni, Alexandra Sasha and Yvon, Fran{\c{c}}ois and Gall{\'e}, Matthias Albert Villanova, del Moral and and Ruwase, Olatunji and Bawden, Rachel and Suarez, Pedro Ortiz and Sanh, Victor and Laurencon, Hugo and Jernite, Yacine and Launay, Julien and Mitchell, Margaret and Raffel, Colin},
  journal={arXiv preprint arXiv:2211.05100},
  year={2022}
}

@article{liu2022dynamic,
  title={Dynamic sparse attention for scalable transformer acceleration},
  author={Liu, Liu and Qu, Zheng and Chen, Zhaodong and Tu, Fengbin and Ding, Yufei and Xie, Yuan},
  journal={IEEE Transactions on Computers},
  volume={71},
  number={12},
  pages={3165--3178},
  year={2022},
  publisher={IEEE}
}

@article{kitaev2020reformer,
  title={{Reformer: The efficient Transformer}},
  author={Kitaev, Nikita and Kaiser, {\L}ukasz and Levskaya, Anselm},
  journal={arXiv preprint arXiv:2001.04451},
  year={2020}
}

@article{dosovitskiy2020image,
  title={{An image is worth 16x16 words: Transformers for image recognition at scale}},
  author={Dosovitskiy, Alexey and Beyer, Lucas and Kolesnikov, Alexander and Weissenborn, Dirk and Zhai, Xiaohua and Unterthiner, Thomas and Dehghani, Mostafa and Minderer, Matthias and Heigold, Georg and Gelly, Sylvain and Uszkoreit, Jakob and Houlsby, Neil},
  journal={arXiv preprint arXiv:2010.11929},
  year={2020}
}

@article{rouhani2023microscaling,
  title={Microscaling data formats for deep learning},
  author={Rouhani, Bita Darvish and Zhao, Ritchie and More, Ankit and Hall, Mathew and Khodamoradi, Alireza and Deng, Summer and Choudhary, Dhruv and Cornea, Marius and Dellinger, Eric and Denolf, Kristof and others},
  journal={arXiv preprint arXiv:2310.10537},
  year={2023}
}

@inproceedings{sheng2023flexgen,
  title={Flexgen: High-throughput generative inference of large language models with a single gpu},
  author={Sheng, Ying and Zheng, Lianmin and Yuan, Binhang and Li, Zhuohan and Ryabinin, Max and Chen, Beidi and Liang, Percy and R{\'e}, Christopher and Stoica, Ion and Zhang, Ce},
  booktitle={International Conference on Machine Learning},
  pages={31094--31116},
  year={2023},
  organization={PMLR}
}

@inproceedings{jang2024figna,
  title={{FIGNA: Integer unit-based accelerator design for FP-INT GEMM preserving numerical accuracy}},
  author={Jang, Jaeyong and Kim, Yulhwa and Lee, Juheun and Kim, Jae-Joon},
  booktitle={2024 IEEE International Symposium on High-Performance Computer Architecture (HPCA)},
  pages={760--773},
  year={2024},
  organization={IEEE}
}

@inproceedings{fang2025anda,
  title={{Anda: Unlocking efficient LLM inference with a variable-length grouped activation data format}},
  author={Fang, Chao and Shi, Man and Geens, Robin and Symons, Arne and Wang, Zhongfeng and Verhelst, Marian},
  booktitle={2025 IEEE International Symposium on High Performance Computer Architecture (HPCA)},
  pages={1467--1481},
  year={2025},
  organization={IEEE}
}

@inproceedings{chen2025bitmod,
  title={{Bitmod: Bit-serial mixture-of-datatype LLM acceleration}},
  author={Chen, Yuzong and AbouElhamayed, Ahmed F and Dai, Xilai and Wang, Yang and Andronic, Marta and Constantinides, George A and Abdelfattah, Mohamed S},
  booktitle={2025 IEEE International Symposium on High Performance Computer Architecture (HPCA)},
  pages={1082--1097},
  year={2025},
  organization={IEEE}
}

@article{tuli2023acceltran,
  title={AccelTran: A sparsity-aware accelerator for dynamic inference with transformers},
  author={Tuli, Shikhar and Jha, Niraj K},
  journal={IEEE Transactions on Computer-Aided Design of Integrated Circuits and Systems},
  volume={42},
  number={11},
  pages={4038--4051},
  year={2023},
  publisher={IEEE}
}

@article{zhao2024hardware,
  title={{Hardware-software co-design enabling static and dynamic sparse attention mechanisms}},
  author={Zhao, Jieru and Zeng, Pai and Shen, Guan and Chen, Quan and Guo, Minyi},
  journal={IEEE Transactions on Computer-Aided Design of Integrated Circuits and Systems},
  year={2024},
  publisher={IEEE}
}

@inproceedings{shen2022salo,
  title={{SALO: An efficient spatial accelerator enabling hybrid sparse attention mechanisms for long sequences}},
  author={Shen, Guan and Zhao, Jieru and Chen, Quan and Leng, Jingwen and Li, Chao and Guo, Minyi},
  booktitle={Proceedings of the 59th ACM/IEEE Design Automation Conference},
  pages={571--576},
  year={2022}
}

@inproceedings{you2023vitcod,
  title={{ViTCoD: Vision Transformer acceleration via dedicated algorithm and accelerator co-design}},
  author={You, Haoran and Sun, Zhanyi and Shi, Huihong and Yu, Zhongzhi and Zhao, Yang and Zhang, Yongan and Li, Chaojian and Li, Baopu and Lin, Yingyan},
  booktitle={Proceedings of the IEEE International Symposium on High-Performance Computer Architecture (HPCA)},
  pages={273--286},
  year={2023}
}

@article{hendrycks2020measuring,
  title={Measuring massive multitask language understanding},
  author={Hendrycks, Dan and Burns, Collin and Basart, Steven and Zou, Andy and Mazeika, Mantas and Song, Dawn and Steinhardt, Jacob},
  journal={arXiv preprint arXiv:2009.03300},
  year={2020}
}

@inproceedings{wang2023cta,
  title={{CTA: Hardware-software co-design for compressed token attention mechanism}},
  author={Wang, Haoran and Xu, Haobo and Wang, Ying and Han, Yinhe},
  booktitle={2023 IEEE International Symposium on High-Performance Computer Architecture (HPCA)},
  pages={429--441},
  year={2023},
  organization={IEEE}
}

@article{austin2021program,
  title={Program synthesis with large language models},
  author={Austin, Jacob and Odena, Augustus and Nye, Maxwell and Bosma, Maarten and Michalewski, Henryk and Dohan, David and Jiang, Ellen and Cai, Carrie and Terry, Michael and Le, Quoc and Sutton Charles},
  journal={arXiv preprint arXiv:2108.07732},
  year={2021}
}

@article{wang2021efficient,
  title={{An efficient stochastic convolution architecture based on fast FIR algorithm}},
  author={Wang, Huizheng and Xu, Weihong and Zhang, Zaichen and You, Xiaohu and Zhang, Chuan},
  journal={IEEE Transactions on Circuits and Systems II: Express Briefs},
  volume={69},
  number={3},
  pages={984--988},
  year={2021},
  publisher={IEEE}
}

@article{tian2023scan,
  title={Scan and snap: Understanding training dynamics and token composition in 1-layer transformer},
  author={Tian, Yuandong and Wang, Yiping and Chen, Beidi and Du, Simon S},
  journal={Advances in neural information processing systems},
  volume={36},
  pages={71911--71947},
  year={2023}
}

@book{brent2010modern,
  title={{Modern computer arithmetic}},
  author={Brent, Richard P and Zimmermann, Paul},
  volume={18},
  year={2010},
  publisher={Cambridge University Press}
}

@article{wang2021efficientMIMO,
  title={{An efficient detector for massive MIMO based on improved matrix partition}},
  author={Wang, Huizheng and Ji, Yahui and Shen, Yifei and Song, Wenqing and Li, Muhao and You, Xiaohu and Zhang, Chuan},
  journal={IEEE Transactions on Signal Processing},
  volume={69},
  pages={2971--2986},
  year={2021},
  publisher={IEEE}
}

@inproceedings{li2020ftrans,
  title={{FTRANS: Energy-efficient acceleration of Transformers using FPGA}},
  author={Li, Bingbing and Pandey, Santosh and Fang, Haowen and Lyv, Yanjun and Li, Ji and Chen, Jieyang and Xie, Mimi and Wan, Lipeng and Liu, Hang and Ding, Caiwen},
  booktitle={Proceedings of the ACM/IEEE International Symposium on Low Power Electronics and Design},
  pages={175--180},
  year={2020}
}

@article{muralimanohar2009cacti,
  title={{CACTI 6.0: A tool to model large caches}},
  author={Muralimanohar, Naveen and Balasubramonian, Rajeev and Jouppi, Norman P},
  journal={HP laboratories},
  volume={27},
  pages={28},
  year={2009}
}

@inproceedings{ham20203,
  title={{A$^3$: Accelerating attention mechanisms in neural networks with approximation}},
  author={Ham, Tae Jun and Jung, Sung Jun and Kim, Seonghak and Oh, Young H and Park, Yeonhong and Song, Yoonho and Park, Jung-Hun and Lee, Sanghee and Park, Kyoung and Lee, Jae W and Jeong, Deog-Kyoon},
  booktitle={Proceedings of the IEEE International Symposium on High Performance Computer Architecture (HPCA)},
  pages={328--341},
  year={2020}
}

@inproceedings{wang2021spatten,
  title={{SpAtten: Efficient sparse attention architecture with cascade token and head pruning}},
  author={Wang, Hanrui and Zhang, Zhekai and Han, Song},
  booktitle={Proceedings of the IEEE International Symposium on High-Performance Computer Architecture (HPCA)},
  pages={97--110},
  year={2021}
}

@inproceedings{qin2023fact,
  title={{Fact: FFN-attention co-optimized Transformer architecture with eager correlation prediction}},
  author={Qin, Yubin and Wang, Yang and Deng, Dazheng and Zhao, Zhiren and Yang, Xiaolong and Liu, Leibo and Wei, Shaojun and Hu, Yang and Yin, Shouyi},
  booktitle={Proceedings of the 50th Annual International Symposium on Computer Architecture},
  pages={1--14},
  year={2023}
}

@inproceedings{lu2021sanger,
  title={{Sanger: A co-design framework for enabling sparse attention using reconfigurable architecture}},
  author={Lu, Liqiang and Jin, Yicheng and Bi, Hangrui and Luo, Zizhang and Li, Peng and Wang, Tao and Liang, Yun},
  booktitle={Proceedings of the 54th Annual IEEE/ACM International Symposium on Microarchitecture},
  pages={977--991},
  year={2021}
}

@inproceedings{wang2024sofa,
  title={{SOFA: A compute-memory optimized sparsity accelerator via cross-stage coordinated tiling}},
  author={Wang, Huizheng and Fang, Jiahao and Tang, Xinru and Yue, Zhiheng and Li, Jinxi and Qin, Yubin and Guan, Sihan and Yang, Qinze and Wang, Yang and Li, Chao and others},
  booktitle={2024 57th IEEE/ACM International Symposium on Microarchitecture (MICRO)},
  pages={1247--1263},
  year={2024},
  organization={IEEE}
}

@inproceedings{delmas2019bit,
  title={{Bit-tactical: A software/hardware approach to exploiting value and bit sparsity in neural networks}},
  author={Delmas Lascorz, Alberto and Judd, Patrick and Stuart, Dylan Malone and Poulos, Zissis and Mahmoud, Mostafa and Sharify, Sayeh and Nikolic, Milos and Siu, Kevin and Moshovos, Andreas},
  booktitle={Proceedings of the Twenty-Fourth International Conference on Architectural Support for Programming Languages and Operating Systems},
  pages={749--763},
  year={2019}
}

@inproceedings{judd2016stripes,
  title={{Stripes: Bit-serial deep neural network computing}},
  author={Judd, Patrick and Albericio, Jorge and Hetherington, Tayler and Aamodt, Tor M and Moshovos, Andreas},
  booktitle={Proceedings of the 49th Annual IEEE/ACM International Symposium on Microarchitecture (MICRO)},
  pages={1--12},
  year={2016}
}

@inproceedings{albericio2017bit,
  title={Bit-pragmatic deep neural network computing},
  author={Albericio, Jorge and Delm{\'a}s, Alberto and Judd, Patrick and Sharify, Sayeh and O'Leary, Gerard and Genov, Roman and Moshovos, Andreas},
  booktitle={Proceedings of the 50th annual IEEE/ACM international symposium on microarchitecture},
  pages={382--394},
  year={2017}
}

@inproceedings{yang2021fusekna,
  title={{FuseKNA: Fused kernel convolution based accelerator for deep neural networks}},
  author={Yang, Jianxun and Zhang, Zhao and Liu, Zhuangzhi and Zhou, Jing and Liu, Leibo and Wei, Shaojun and Yin, Shouyi},
  booktitle={Proceddings of the IEEE International Symposium on High-Performance Computer Architecture (HPCA)},
  pages={894--907},
  year={2021}
}

@inproceedings{han2024bitnn,
  title={{BitNN: A bit-serial accelerator for k-nearest neighbor search in point clouds}},
  author={Han, Meng and Wang, Liang and Xiao, Limin and Zhang, Hao and Cai, Tianhao and Xu, Jiale and Wu, Yibo and Zhang, Chenhao and Xu, Xiangrong},
  booktitle={Proceddings of the ACM/IEEE 51st Annual International Symposium on Computer Architecture (ISCA)},
  pages={1278--1292},
  year={2024}
}

@article{zhou2022energon,
  title={{Energon: Toward efficient acceleration of Transformers using dynamic sparse attention}},
  author={Zhou, Zhe and Liu, Junlin and Gu, Zhenyu and Sun, Guangyu},
  journal={IEEE Transactions on Computer-Aided Design of Integrated Circuits and Systems},
  volume={42},
  number={1},
  pages={136--149},
  year={2022},
  publisher={IEEE}
}

@inproceedings{chang2016understanding,
  title={Understanding latency variation in modern DRAM chips: Experimental characterization, analysis, and optimization},
  author={Chang, Kevin K and Kashyap, Abhijith and Hassan, Hasan and Ghose, Saugata and Hsieh, Kevin and Lee, Donghyuk and Li, Tianshi and Pekhimenko, Gennady and Khan, Samira and Mutlu, Onur},
  booktitle={Proceedings of the 2016 ACM SIGMETRICS International Conference on Measurement and Modeling of Computer Science},
  pages={323--336},
  year={2016}
}

@standard{JEDEC_JESD235D,
  author      = {{JEDEC}},
  title       = {{JESD235D: High Bandwidth Memory DRAM (HBM1, HBM2)}},
  type        = {Standard},
  number      = {JESD235D},
  institution = {JEDEC},
  year        = {2021}
}

@inproceedings{sun2017revisiting,
  title={Revisiting unreasonable effectiveness of data in deep learning era},
  author={Sun, Chen and Shrivastava, Abhinav and Singh, Saurabh and Gupta, Abhinav},
  booktitle={Proceedings of the IEEE international conference on computer vision},
  pages={843--852},
  year={2017}
}

@inproceedings{merity2016pointer,
  title={{Pointer sentinel mixture models}},
  author={Merity, Stephen and Xiong, Caiming and Bradbury, James and Socher, Richard},
  booktitle={Proceedings of the International Conference on Learning Representations},
  year={2016}
}

@inproceedings{deng2009imagenet,
  title={{Imagenet: A large-scale hierarchical image database}},
  author={Deng, Jia and Dong, Wei and Socher, Richard and Li, Li-Jia and Li, Kai and Fei-Fei, Li},
  booktitle={Proceedings of the IEEE Conference on Computer Vision and Pattern Recognition},
  pages={248--255},
  year={2009}
}

@inproceedings{wolf2020transformers,
  title={{Transformers: State-of-the-art natural language processing}},
  author={Wolf, Thomas and Debut, Lysandre and Sanh, Victor and Chaumond, Julien and Delangue, Clement and Moi, Anthony and Cistac, Pierric and Rault, Tim and Louf, R{\'e}mi and Funtowicz, Morgan and Davison, Joe and Shleifer, Sam and Platen, Patrick von and Ma, Clara and Jernite, Yacine and Plu, Julien and Xu, Canwen and Scao, Teven Le and Gugger, Sylvain and Drame, Mariama and Lhoest, Quentin and Rush, Alexander},
  booktitle={Proceedings of the Conference on Empirical Methods in Natural Language Processing: System Demonstrations},
  pages={38--45},
  year={2020}
}

@article{paszke2017automatic,
  title={{Automatic differentiation in PyTorch}},
  author={Paszke, Adam and Gross, Sam and Chintala, Soumith and Chanan, Gregory and Yang, Edward and DeVito, Zachary and Lin, Zeming and Desmaison, Alban and Antiga, Luca and Lerer, Adam},
  year={2017}
}

@article{touvron2023llama2,
  title={{Llama 2: Open foundation and fine-tuned chat models}},
  author={Touvron, Hugo and Martin, Louis and Stone, Kevin and Albert, Peter and Almahairi, Amjad and Babaei, Yasmine and Bashlykov, Nikolay and Batra, Soumya and Bhargava, Prajjwal and Bhosale, Shruti and Dan, Bikel and Lukas, Blecher and Cristian, Canton Ferrer and Moya, Chen and Guillem, Cucurull and David, Esiobu and Jude, Fernandes and  Jeremy, Fu and Wenyin, Fu and Brian, Fuller and Cynthia, Gao and Vedanuj, Goswami and Naman, Goyal and Anthony, Hartshorn and Saghar, Hosseini and Rui, Hou and Hakan, Inan and Marcin, Kardas},
  journal={arXiv preprint arXiv:2307.09288},
  year={2023}
}

@article{lin2023videodirectorgpt,
  title={{Videodirectorgpt: Consistent multi-scene video generation via LLM-guided planning}},
  author={Lin, Han and Zala, Abhay and Cho, Jaemin and Bansal, Mohit},
  journal={arXiv preprint arXiv:2309.15091},
  year={2023}
}

@article{zhang2022opt,
  title={{OPT: Open pre-trained transformer language models}},
  author={Zhang, Susan and Roller, Stephen and Goyal, Naman and Artetxe, Mikel and Chen, Moya and Chen, Shuohui and Dewan, Christopher and Diab, Mona and Li, Xian and Lin, Xi Victoria and Mihaylov, Todor and Ott, Myle and Shleifer, Sam and Shuster, Kurt and Simig, Daniel and Singh Koura, Punit and Sridhar, Anjali and Wang, Tianlu and Zettlemoyer, Luke},
  journal={arXiv preprint arXiv:2205.01068},
  year={2022}
}

@inproceedings{park2024token,
  title={Token-Picker: Accelerating Attention in Text Generation with Minimized Memory Transfer via Probability Estimation},
  author={Park, Junyoung and Kang, Myeonggu and Han, Yunki and Kim, Yang-Gon and Shin, Jaekang and Kim, Lee-Sup},
  booktitle={Proceedings of the 61st ACM/IEEE Design Automation Conference},
  pages={1--6},
  year={2024}
}

@article{lee2025clat,
  title={CLAT: A Clustering-Based Attention Transformer Accelerator for Low-Latency Text Generation in LLMs},
  author={Lee, Sunwoo and Kim, Beomseok and Park, Jeongwoo and Jeon, Dongsuk},
  journal={IEEE Transactions on Circuits and Systems I: Regular Papers},
  year={2025},
  publisher={IEEE}
}

@inproceedings{cho2024sparc,
  title={SpARC: Token Similarity-Aware Sparse Attention Transformer Accelerator via Row-wise Clustering},
  author={Cho, Han and Kim, Dongjun and Hwang, Seung-Eon and Park, Jongsun},
  booktitle={Proceedings of the 61st ACM/IEEE Design Automation Conference},
  pages={1--6},
  year={2024}
}

@inproceedings{song2024tsacc,
  title={TSAcc: An Efficient T empo-S patial Similarity Aware Acc elerator for Attention Acceleration},
  author={Song, Zhuoran and Qi, Chunyu and Yao, Yuanzheng and Zhou, Peng and Zi, Yanyi and Wang, Nan and Liang, Xiaoyao},
  booktitle={Proceedings of the 61st ACM/IEEE Design Automation Conference},
  pages={1--6},
  year={2024}
}

@article{kachris2025survey,
  title={A survey on hardware accelerators for large language models},
  author={Kachris, Christoforos},
  journal={Applied Sciences},
  volume={15},
  number={2},
  pages={586},
  year={2025},
  publisher={MDPI}
}

@article{jiang2024minference,
  title={{MInference 1.0: Accelerating prefilling for long-context LLMs via dynamic sparse attention}},
  author={Jiang, Huiqiang and Li, Yucheng and Zhang, Chengruidong and Wu, Qianhui and Luo, Xufang and Ahn, Surin and Han, Zhenhua and Abdi, Amir H and Li, Dongsheng and Lin, Chin-Yew and Yang, Yuqing and Qiu Lili},
  journal={arXiv preprint arXiv:2407.02490},
  year={2024}
}

@article{correia2019adaptively,
  title={{Adaptively sparse Transformers}},
  author={Correia, Gon{\c{c}}alo M and Niculae, Vlad and Martins, Andr{\'e} FT},
  journal={arXiv preprint arXiv:1909.00015},
  year={2019}
}

@inproceedings{ham2021elsa,
  title={{ELSA: Hardware-software co-design for efficient, lightweight self-attention mechanism in neural networks}},
  author={Ham, Tae Jun and Lee, Yejin and Seo, Seong Hoon and Kim, Soosung and Choi, Hyunji and Jung, Sung Jun and Lee, Jae W},
  booktitle={Prceedings of the 48th ACM/IEEE Annual International Symposium on Computer Architecture (ISCA)},
  pages={692--705},
  year={2021}
}

@article{xiao2023efficient,
  title={Efficient streaming language models with attention sinks},
  author={Xiao, Guangxuan and Tian, Yuandong and Chen, Beidi and Han, Song and Lewis, Mike},
  journal={arXiv preprint arXiv:2309.17453},
  year={2023}
}

@article{beltagy2020longformer,
  title={{Longformer: The long-document Transformer}},
  author={Beltagy, Iz and Peters, Matthew E and Cohan, Arman},
  journal={arXiv preprint arXiv:2004.05150},
  year={2020}
}

@article{zaheer2020big,
  title={{Big bird: Transformers for longer sequences}},
  author={Zaheer, Manzil and Guruganesh, Guru and Dubey, Kumar Avinava and Ainslie, Joshua and Alberti, Chris and Ontanon, Santiago and Pham, Philip and Ravula, Anirudh and Wang, Qifan and Yang, Li and Ahmed, Amr},
  journal={Advances in neural information processing systems},
  volume={33},
  pages={17283--17297},
  year={2020}
}

@article{wang2022energy,
  title={{An energy-efficient Transformer processor exploiting dynamic weak relevances in global attention}},
  author={Wang, Yang and Qin, Yubin and Deng, Dazheng and Wei, Jingchuan and Zhou, Yang and Fan, Yuanqi and Chen, Tianbao and Sun, Hao and Liu, Leibo and Wei, Shaojun and Yin, Shouyi},
  journal={IEEE Journal of Solid-State Circuits},
  volume={58},
  number={1},
  pages={227--242},
  year={2022},
  publisher={IEEE}
}

@article{wang2023efficient,
  title={{An efficient approximate expectation propagation detector with block-diagonal Neumann-series}},
  author={Wang, Huizheng and Cheng, Bingyang and Tan, Xiaosi and You, Xiaohu and Zhang, Chuan},
  journal={IEEE Transactions on Circuits and Systems I: Regular Papers},
  volume={70},
  number={3},
  pages={1403--1416},
  year={2023},
  publisher={IEEE}
}

@article{yang2022dtatrans,
  title={{DTATrans: Leveraging dynamic token-based quantization with accuracy compensation mechanism for efficient Transformer architecture}},
  author={Yang, Tao and Ma, Fei and Li, Xiaoling and Liu, Fangxin and Zhao, Yilong and He, Zhezhi and Jiang, Li},
  journal={IEEE Transactions on Computer-Aided Design of Integrated Circuits and Systems},
  volume={42},
  number={2},
  pages={509--520},
  year={2022},
  publisher={IEEE}
}

@inproceedings{qu2022dota,
  title={{DOTA: Detect and omit weak attentions for scalable Transformer acceleration}},
  author={Qu, Zheng and Liu, Liu and Tu, Fengbin and Chen, Zhaodong and Ding, Yufei and Xie, Yuan},
  booktitle={Proceedings of the 27th ACM International Conference on Architectural Support for Programming Languages and Operating Systems},
  pages={14--26},
  year={2022}
}

@article{conover2023free,
  title={{Free dolly: Introducing the world’s first truly open instruction-tuned LLM}},
  author={Conover, Mike and Hayes, Matt and Mathur, Ankit and Xie, Jianwei and Wan, Jun and Shah, Sam and Ghodsi, Ali and Wendell, Patrick and Zaharia, Matei and Xin, Reynold},
  journal={Company Blog of Databricks},
  year={2023}
}

@inproceedings{fan2022adaptable,
  title={{Adaptable butterfly accelerator for attention-based NNs via hardware and algorithm co-design}},
  author={Fan, Hongxiang and Chau, Thomas and Venieris, Stylianos I and Lee, Royson and Kouris, Alexandros and Luk, Wayne and Lane, Nicholas D and Abdelfattah, Mohamed S},
  booktitle={Proceedings of the 55th IEEE/ACM International Symposium on Microarchitecture (MICRO)},
  pages={599--615},
  year={2022}
}

@inproceedings{dass2023vitality,
  title={Vitality: Unifying low-rank and sparse approximation for vision transformer acceleration with a linear taylor attention},
  author={Dass, Jyotikrishna and Wu, Shang and Shi, Huihong and Li, Chaojian and Ye, Zhifan and Wang, Zhongfeng and Lin, Yingyan},
  booktitle={2023 IEEE International Symposium on High-Performance Computer Architecture (HPCA)},
  pages={415--428},
  year={2023},
  organization={IEEE}
}

@article{fang2022algorithm,
  title={{An algorithm-hardware co-optimized framework for accelerating N:M sparse Transformers}},
  author={Fang, Chao and Zhou, Aojun and Wang, Zhongfeng},
  journal={IEEE Transactions on Very Large Scale Integration (VLSI) Systems},
  volume={30},
  number={11},
  pages={1573--1586},
  year={2022},
  publisher={IEEE}
}

@inproceedings{hanson2022cascading,
  title={{Cascading structured pruning: Enabling high data reuse for sparse DNN accelerators}},
  author={Hanson, Edward and Li, Shiyu and Li, Hai'Helen' and Chen, Yiran},
  booktitle={Proceedings of the 49th Annual International Symposium on Computer Architecture},
  pages={522--535},
  year={2022}
}

@inproceedings{kanellopoulos2019smash,
  title={{SMASH: Co-designing software compression and hardware-accelerated indexing for efficient sparse matrix operations}},
  author={Kanellopoulos, Konstantinos and Vijaykumar, Nandita and Giannoula, Christina and Azizi, Roknoddin and Koppula, Skanda and Ghiasi, Nika Mansouri and Shahroodi, Taha and Luna, Juan Gomez and Mutlu, Onur},
  booktitle={Proceedings of the 52nd annual IEEE/ACM international symposium on microarchitecture},
  pages={600--614},
  year={2019}
}

@inproceedings{liu2022s2ta,
  title={{S2TA: Exploiting structured sparsity for energy-efficient mobile CNN acceleration}},
  author={Liu, Zhi-Gang and Whatmough, Paul N and Zhu, Yuhao and Mattina, Matthew},
  booktitle={Proceedings of the IEEE International Symposium on High-Performance Computer Architecture (HPCA)},
  pages={573--586},
  year={2022}
}

@inproceedings{mahmoud2020tensordash,
  title={{TensorDash: Exploiting sparsity to accelerate deep neural network training}},
  author={Mahmoud, Mostafa and Edo, Isak and Zadeh, Ali Hadi and Awad, Omar Mohamed and Pekhimenko, Gennady and Albericio, Jorge and Moshovos, Andreas},
  booktitle={Proceedings of the 53rd Annual IEEE/ACM International Symposium on Microarchitecture (MICRO)},
  pages={781--795},
  year={2020}
}

@inproceedings{yang2020procrustes,
  title={{Procrustes: A dataflow and accelerator for sparse deep neural network training}},
  author={Yang, Dingqing and Ghasemazar, Amin and Ren, Xiaowei and Golub, Maximilian and Lemieux, Guy and Lis, Mieszko},
  booktitle={Proceedings of the 53rd Annual IEEE/ACM International Symposium on Microarchitecture (MICRO)},
  pages={711--724},
  year={2020}
}

@article{walia2021fast,
  title={{Fast and low-power quantized fixed posit high-accuracy DNN implementation}},
  author={Walia, Sumit and Tej, Bachu Varun and Kabra, Arpita and Devnath, Joydeep and Mekie, Joycee},
  journal={IEEE Transactions on Very Large Scale Integration (VLSI) Systems},
  volume={30},
  number={1},
  pages={108--111},
  year={2021},
  publisher={IEEE}
}

@inproceedings{srivastava2020matraptor,
  title={{MatRaptor: A sparse-sparse matrix multiplication accelerator based on row-wise product}},
  author={Srivastava, Nitish and Jin, Hanchen and Liu, Jie and Albonesi, David and Zhang, Zhiru},
  booktitle={Proceedings of the 53rd Annual IEEE/ACM International Symposium on Microarchitecture (MICRO)},
  pages={766--780},
  year={2020}
}

@inproceedings{shin2022griffin,
  title={{Griffin: Rethinking sparse optimization for deep learning architectures}},
  author={Shin, Jong Hoon and Shafiee, Ali and Pedram, Ardavan and Abdel-Aziz, Hamzah and Li, Ling and Hassoun, Joseph},
  booktitle={Proceedings of the IEEE International Symposium on High-Performance Computer Architecture (HPCA)},
  pages={861--875},
  year={2022}
}

@inproceedings{sadi2019efficient,
  title={{Efficient SPMV operation for large and highly sparse matrices using scalable multi-way merge parallelization}},
  author={Sadi, Fazle and Sweeney, Joe and Low, Tze Meng and Hoe, James C and Pileggi, Larry and Franchetti, Franz},
  booktitle={Proceedings of the 52nd Annual IEEE/ACM International Symposium on Microarchitecture},
  pages={347--358},
  year={2019}
}

@inproceedings{wang2025mcbp,
  title={{MCBP: A memory-compute efficient LLM inference accelerator leveraging bit-slice-enabled sparsity and repetitiveness}},
  author={Wang, Huizheng and Wang, Zichuan and Yue, Zhiheng and Long, Yousheng and Wei, Taiquan and Yang, Jianxun and Wang, Yang and Li, Chao and Wei, Shaojun and Hu, Yang and others},
  booktitle={Proceedings of the 58th IEEE/ACM International Symposium on Microarchitecture},
  pages={1592--1608},
  year={2025}
}

@inproceedings{rucker2021capstan,
  title={{Capstan: A vector RDA for sparsity}},
  author={Rucker, Alexander and Vilim, Matthew and Zhao, Tian and Zhang, Yaqi and Prabhakar, Raghu and Olukotun, Kunle},
  booktitle={Proceedings of the 54th Annual IEEE/ACM International Symposium on Microarchitecture},
  pages={1022--1035},
  year={2021}
}

@inproceedings{wang2018low,
  title={{Low-complexity Winograd convolution architecture based on stochastic computing}},
  author={Wang, Huizheng and Zhang, Zaichen and You, Xiaohu and Zhang, Chuan},
  booktitle={2018 IEEE 23rd International Conference on Digital Signal Processing (DSP)},
  pages={1--5},
  year={2018},
  organization={IEEE}
}

@article{wang2025bitstopper,
  title={{BitStopper: An efficient Transformer attention accelerator via stage-fusion and early termination}},
  author={Wang, Huizheng and Wang, Hongbin and Wei, Shaojun and Hu, Yang and Yin, Shouyi},
  journal={arXiv preprint arXiv:2512.06457},
  year={2025}
}

@article{wang2025lapa,
  title={{LAPA: Log-domain prediction-driven dynamic sparsity accelerator for Transformer model}},
  author={Wang, Huizheng and Wang, Hongbin and Wei, Shaojun and Hu, Yang and Yin, Shouyi},
  journal={arXiv preprint arXiv:2512.07855},
  year={2025}
}

@inproceedings{pavon2021via,
  title={{VIA: A smart scratchpad for vector units with application to sparse matrix computations}},
  author={Pavon, Julian and Valdivieso, Ivan Vargas and Barredo, Adrian and Marimon, Joan and Moreto, Miquel and Moll, Francesc and Unsal, Osman and Valero, Mateo and Cristal, Adrian},
  booktitle={Proceedings of the IEEE International Symposium on High-Performance Computer Architecture (HPCA)},
  pages={921--934},
  year={2021}
}

@inproceedings{li2022accelerating,
  title={{Accelerating attention through gradient-based learned runtime pruning}},
  author={Li, Zheng and Ghodrati, Soroush and Yazdanbakhsh, Amir and Esmaeilzadeh, Hadi and Kang, Mingu},
  booktitle={Proceedings of the 49th Annual International Symposium on Computer Architecture},
  pages={902--915},
  year={2022}
}

@article{grattafiori2024llama,
  title={The llama 3 herd of models},
  author={Grattafiori, Aaron and Dubey, Abhimanyu and Jauhri, Abhinav and Pandey, Abhinav and Kadian, Abhishek and Al-Dahle, Ahmad and Letman, Aiesha and Mathur, Akhil and Schelten, Alan and Vaughan, Alex and Yang, Amy and Fan, Angela and Goyal, Anirudh and Hartshorn, Anthony and Yang, Aobo and Mitra, Archi and Sravankumar, Archie and Korenev, Artem and Hinsvark, Arthur and Rao, Arun and Zhang, Aston and Rodriguez, Aurelien and Gregerson, Austen and Spataru, Ava and Roziere, Baptiste and Biron, Bethany and Tang, Binh and Chern, Bobbie and Caucheteux, Charlotte and Nayak, Chaya and Bi, Chloe and Marra, Chris and McConnell, Chris and Keller, Christian and Touret, Christophe and Wu, Chunyang and Wong, Corinne and Canton Ferrer, Cristian},
  journal={arXiv preprint arXiv:2407.21783},
  year={2024}
}

@inproceedings{li2021escalate,
  title={{ESCALATE: Boosting the efficiency of sparse CNN accelerator with kernel decomposition}},
  author={Li, Shiyu and Hanson, Edward and Qian, Xuehai and Li, Hai" Helen" and Chen, Yiran},
  booktitle={Proceedings of the 54th Annual IEEE/ACM International Symposium on Microarchitecture},
  pages={992--1004},
  year={2021}
}

@inproceedings{li2022ristretto,
  title={{Ristretto: An atomized processing architecture for sparsity-condensed stream flow in CNN}},
  author={Li, Gang and Xu, Weixiang and Song, Zhuoran and Jing, Naifeng and Cheng, Jian and Liang, Xiaoyao},
  booktitle={Proceedings of the 55th IEEE/ACM International Symposium on Microarchitecture (MICRO)},
  pages={1434--1450},
  year={2022}
}

@inproceedings{xiao2023smoothquant,
  title={{Smoothquant: Accurate and efficient post-training quantization for large language models}},
  author={Xiao, Guangxuan and Lin, Ji and Seznec, Mickael and Wu, Hao and Demouth, Julien and Han, Song},
  booktitle={International Conference on Machine Learning},
  pages={38087--38099},
  year={2023},
  organization={PMLR}
}

@inproceedings{lew2022anticipating,
  title={{Anticipating and eliminating redundant computations in accelerated sparse training}},
  author={Lew, Jonathan S and Liu, Yunpeng and Gong, Wenyi and Goli, Negar and Evans, R David and Aamodt, Tor M},
  booktitle={Proceedings of the 49th Annual International Symposium on Computer Architecture},
  pages={536--551},
  year={2022}
}

@inproceedings{hojabr2021spaghetti,
  title={{SPAGHETTI: Streaming accelerators for highly sparse GEMM on FPGAs}},
  author={Hojabr, Reza and Sedaghati, Ali and Sharifian, Amirali and Khonsari, Ahmad and Shriraman, Arrvindh},
  booktitle={Proceedings of the IEEE International Symposium on High-Performance Computer Architecture (HPCA)},
  pages={84--96},
  year={2021}
}

@inproceedings{hegde2019extensor,
  title={{ExTensor: An accelerator for sparse tensor algebra}},
  author={Hegde, Kartik and Asghari-Moghaddam, Hadi and Pellauer, Michael and Crago, Neal and Jaleel, Aamer and Solomonik, Edgar and Emer, Joel and Fletcher, Christopher W},
  booktitle={Proceedings of the 52nd Annual IEEE/ACM International Symposium on Microarchitecture},
  pages={319--333},
  year={2019}
}

@inproceedings{gudaparthi2022candles,
  title={{CANDLES: Channel-aware novel dataflow-microarchitecture co-design for low energy sparse neural network acceleration}},
  author={Gudaparthi, Sumanth and Singh, Sarabjeet and Narayanan, Surya and Balasubramonian, Rajeev and Sathe, Visvesh},
  booktitle={Proceedings of the IEEE International Symposium on high-performance computer architecture (HPCA)},
  pages={876--891},
  year={2022}
}

@inproceedings{gondimalla2019sparten,
  title={{SparTen: A sparse tensor accelerator for convolutional neural networks}},
  author={Gondimalla, Ashish and Chesnut, Noah and Thottethodi, Mithuna and Vijaykumar, TN},
  booktitle={Proceedings of the 52nd Annual IEEE/ACM International Symposium on Microarchitecture},
  pages={151--165},
  year={2019}
}

@inproceedings{hong2022dfx,
  title={{DFX: A low-latency multi-FPGA appliance for accelerating Transformer-based text generation}},
  author={Hong, Seongmin and Moon, Seungjae and Kim, Junsoo and Lee, Sungjae and Kim, Minsub and Lee, Dongsoo and Kim, Joo-Young},
  booktitle={Proceedings of the 55th IEEE/ACM International Symposium on Microarchitecture (MICRO)},
  pages={616--630},
  year={2022}
}

@article{delmas2017dynamic,
  title={{Dynamic stripes: Exploiting the dynamic precision requirements of activation values in neural networks}},
  author={Delmas, Alberto and Judd, Patrick and Sharify, Sayeh and Moshovos, Andreas},
  journal={arXiv preprint arXiv:1706.00504},
  year={2017}
}

@inproceedings{yazdanbakhsh2022sparse,
  title={{Sparse attention acceleration with synergistic in-memory pruning and on-chip recomputation}},
  author={Yazdanbakhsh, Amir and Moradifirouzabadi, Ashkan and Li, Zheng and Kang, Mingu},
  booktitle={Proceedings of the 55th IEEE/ACM International Symposium on Microarchitecture (MICRO)},
  pages={744--762},
  year={2022}
}

@article{kim2015ramulator,
  title={{Ramulator: A fast and extensible DRAM simulator}},
  author={Kim, Yoongu and Yang, Weikun and Mutlu, Onur},
  journal={IEEE Computer architecture letters},
  volume={15},
  number={1},
  pages={45--49},
  year={2015},
  publisher={IEEE}
}

@inproceedings{andri2016yodann,
  title={{YodaNN: An ultra-low power convolutional neural network accelerator based on binary weights}},
  author={Andri, Renzo and Cavigelli, Lukas and Rossi, Davide and Benini, Luca},
  booktitle={Proceedings of the IEEE Computer Society Annual Symposium on VLSI (ISVLSI)},
  pages={236--241},
  year={2016}
}

@article{wang2017energy,
  title={{An energy-efficient architecture for binary weight convolutional neural networks}},
  author={Wang, Yizhi and Lin, Jun and Wang, Zhongfeng},
  journal={IEEE Transactions on Very Large Scale Integration (VLSI) Systems},
  volume={26},
  number={2},
  pages={280--293},
  year={2017},
  publisher={IEEE}
}

@inproceedings{kam2025panacea,
  title={{Panacea: Novel DNN accelerator using accuracy-preserving asymmetric quantization and energy-saving bit-slice sparsity}},
  author={Kam, Dongyun and Yun, Myeongji and Yoo, Sunwoo and Hong, Seungwoo and Zhang, Zhengya and Lee, Youngjoo},
  booktitle={Proc. IEEE Int. Symp. High Perform. Comput. Archit. (HPCA)},
  pages={701--715},
  year={2025}
}

@inproceedings{lu2021distilling,
  title={Distilling bit-level sparsity parallelism for general purpose deep learning acceleration},
  author={Lu, Hang and Chang, Liang and Li, Chenglong and Zhu, Zixuan and Lu, Shengjian and Liu, Yanhuan and Zhang, Mingzhe},
  booktitle={MICRO-54: 54th Annual IEEE/ACM International Symposium on Microarchitecture},
  pages={963--976},
  year={2021}
}

@inproceedings{sharify2019laconic,
  title={Laconic deep learning inference acceleration},
  author={Sharify, Sayeh and Lascorz, Alberto Delmas and Mahmoud, Mostafa and Nikolic, Milos and Siu, Kevin and Stuart, Dylan Malone and Poulos, Zissis and Moshovos, Andreas},
  booktitle={Proceedings of the 46th International Symposium on Computer Architecture},
  pages={304--317},
  year={2019}
}

@inproceedings{im2023sibia,
  title={{Sibia: Signed bit-slice architecture for dense DNN acceleration with slice-level sparsity exploitation}},
  author={Im, Dongseok and Park, Gwangtae and Li, Zhiyong and Ryu, Junha and Yoo, Hoi-Jun},
  booktitle={2023 IEEE International Symposium on High-Performance Computer Architecture (HPCA)},
  pages={69--80},
  year={2023},
  organization={IEEE}
}

@misc{Nvidia2023,
  author = {Nvidia},
  title = {{TensorRT-LLM}},
  year = {2023},
  note = {\url{https://github.com/NVIDIA/TensorRT-LLM?tab=readme-ov-file}},
}

@article{bai2023qwen,
  title={Qwen technical report},
  author={Bai, Jinze and Bai, Shuai and Chu, Yunfei and Cui, Zeyu and Dang, Kai and Deng, Xiaodong and Fan, Yang and Ge, Wenbin and Han, Yu and Huang, Fei and Hui, Bingyuan and Ji, Luo and Li, Mei and Lin, Junyang and Lin, Runji and Liu, Dayiheng and Liu, Gao and Lu, Chengqiang and Lu, Keming and Ma, Jianxin and Men, Rui and Ren, Xingzhang and Ren, Xuancheng and Tan, Chuanqi and Tan, Sinan and Tu, Jianhong and Wang, Peng and Wang, Shijie and Wang, Wei and Wu, Shengguang and XU, Benfeng and Xu, Jin and Yang, An and Yang, Hao and Yang, Jian and Yang, Shusheng and Yao, Yang and Yu, Bowen and Yuan, Hongyi and Yuan, Zheng and Zhang, Jianwei and Zhang, Xingxuan and Zhang, Yichang and Zhang, Zhenru and Zhou, Chen and Zhou, Jingren and Zhou, Xiaohuan and Zhu, Tianhang},
  journal={arXiv preprint arXiv:2309.16609},
  year={2023}
}

@inproceedings{deng2021gospa,
  title={{GoSPA: An energy-efficient high-performance globally optimized sparse convolutional neural network accelerator}},
  author={Deng, Chunhua and Sui, Yang and Liao, Siyu and Qian, Xuehai and Yuan, Bo},
  booktitle={Proceedings of the ACM/IEEE 48th Annual International Symposium on Computer Architecture (ISCA)},
  pages={1110--1123},
  year={2021}
}

@inproceedings{asgari2020alrescha,
  title={{ALRESCHA: A lightweight reconfigurable sparse-computation accelerator}},
  author={Asgari, Bahar and Hadidi, Ramyad and Krishna, Tushar and Kim, Hyesoon and Yalamanchili, Sudhakar},
  booktitle={Proceedings of the IEEE International Symposium on High Performance Computer Architecture (HPCA)},
  pages={249--260},
  year={2020}
}

@inproceedings{lo2023bit,
  title={{Bit-serial cache: Exploiting input bit vector repetition to accelerate bit-serial inference}},
  author={Lo, Yun-Chen and Liu, Ren-Shuo},
  booktitle={Proceedings of the 60th ACM/IEEE Design Automation Conference (DAC)},
  pages={1--6},
  year={2023}
}

@inproceedings{yang2023isosceles,
  title={{ISOSceles: Accelerating sparse CNNs through inter-layer pipelining}},
  author={Yang, Yifan and Emer, Joel S and Sanchez, Daniel},
  booktitle={2023 IEEE International Symposium on High-Performance Computer Architecture (HPCA)},
  pages={598--610},
  year={2023},
  organization={IEEE}
}

@inproceedings{guo2025transitive,
  title={{Transitive array: An efficient GEMM accelerator with result reuse}},
  author={Guo, Cong and Wei, Chiyue and Tang, Jiaming and Duan, Bowen and Han, Song and Li, Hai and Chen, Yiran},
  booktitle={Proc. IEEE Annu. Int. Symp. Comput. Archit. (ISCA)},
  pages={990--1004},
  year={2025}
}

@inproceedings{chen2024bbs,
  title={BBS: Bi-directional bit-level sparsity for deep learning acceleration},
  author={Chen, Yuzong and Meng, Jian and Seo, Jae-sun and Abdelfattah, Mohamed S},
  booktitle={2024 57th IEEE/ACM International Symposium on Microarchitecture (MICRO)},
  pages={551--564},
  year={2024},
  organization={IEEE}
}

@inproceedings{qin2020sigma,
  title={{Sigma: A sparse and irregular GEMM accelerator with flexible interconnects for DNN training}},
  author={Qin, Eric and Samajdar, Ananda and Kwon, Hyoukjun and Nadella, Vineet and Srinivasan, Sudarshan and Das, Dipankar and Kaul, Bharat and Krishna, Tushar},
  booktitle={Proceedings of the IEEE International Symposium on High Performance Computer Architecture (HPCA)},
  pages={58--70},
  year={2020}
}

@inproceedings{lee2018unpu,
  title={{UNPU: A 50.6 TOPS/W unified deep neural network accelerator with 1b-to-16b fully-variable weight bit-precision}},
  author={Lee, Jinmook and Kim, Changhyeon and Kang, Sanghoon and Shin, Dongjoo and Kim, Sangyeob and Yoo, Hoi-Jun},
  booktitle={Proceedings of IEEE International Solid-State Circuits Conference-(ISSCC)},
  pages={218--220},
  year={2018}
}

@article{parashar2017scnn,
  title={{SCNN: An accelerator for compressed-sparse convolutional neural networks}},
  author={Parashar, Angshuman and Rhu, Minsoo and Mukkara, Anurag and Puglielli, Antonio and Venkatesan, Rangharajan and Khailany, Brucek and Emer, Joel and Keckler, Stephen W and Dally, William J},
  journal={ACM SIGARCH computer architecture news},
  volume={45},
  number={2},
  pages={27--40},
  year={2017},
  publisher={ACM New York, NY, USA}
}

@article{li2025lut,
  title={LUT-DLA: Lookup Table as Efficient Extreme Low-Bit Deep Learning Accelerator},
  author={Li, Guoyu and Ye, Shengyu and Chen, Chunyun and Wang, Yang and Yang, Fan and Cao, Ting and Liu, Cheng and Sabry, Mohamed M and Yang, Mao},
  journal={arXiv preprint arXiv:2501.10658},
  year={2025}
}

@article{chen2020tpspmv,
  title={{tpSpMV: A two-phase large-scale sparse matrix-vector multiplication kernel for manycore architectures}},
  author={Chen, Yuedan and Xiao, Guoqing and Wu, Fan and Tang, Zhuo and Li, Keqin},
  journal={Information Sciences},
  volume={523},
  pages={279--295},
  year={2020},
  publisher={Elsevier}
}

@inproceedings{kwon2019tensordimm,
  title={{TensorDIMM: A practical near-memory processing architecture for embeddings and tensor operations in deep learning}},
  author={Kwon, Youngeun and Lee, Yunjae and Rhu, Minsoo},
  booktitle={Proceedings of the 52nd Annual IEEE/ACM International Symposium on Microarchitecture},
  pages={740--753},
  year={2019}
}

@inproceedings{Gon2023Eureka,
title = {{Eureka: Efficient tensor cores for one-sided unstructured sparsity in DNN inference}},
author = {Gondimalla, Ashish and Thottethodi, Mithuna and Vijaykumar, T. N.},
booktitle = {Proceedings of the 56th Annual IEEE/ACM International Symposium on Microarchitecture},
pages = {324–337},
year = {2023}
}

@inproceedings{wu2022sparseloop,
  title={{Sparseloop: An analytical approach to sparse tensor accelerator modeling}},
  author={Wu, Yannan Nellie and Tsai, Po-An and Parashar, Angshuman and Sze, Vivienne and Emer, Joel S},
  booktitle={Proceedings of the 55th Annual IEEE/ACM International Symposium on Microarchitecture (MICRO)},
  pages={1377--1395},
  year={2022}
}

@inproceedings{wu2023highlight,
  title={{HighLight: Efficient and flexible DNN acceleration with hierarchical structured sparsity}},
  author={Wu, Yannan Nellie and Tsai, Po-An and Muralidharan, Saurav and Parashar, Angshuman and Sze, Vivienne and Emer, Joel},
  booktitle={Proceedings of the 56th Annual IEEE/ACM International Symposium on Microarchitecture},
  pages={1106--1120},
  year={2023}
}

@inproceedings{huang2023rm,
  title={{RM-STC: Row-merge dataflow inspired GPU sparse tensor core for energy-efficient sparse acceleration}},
  author={Huang, Guyue and Wang, Zhengyang and Tsai, Po-An and Zhang, Chen and Ding, Yufei and Xie, Yuan},
  booktitle={Proceedings of the 56th Annual IEEE/ACM International Symposium on Microarchitecture},
  pages={338--352},
  year={2023}
}

@inproceedings{wang2021dual,
  title={{Dual-side sparse tensor core}},
  author={Wang, Yang and Zhang, Chen and Xie, Zhiqiang and Guo, Cong and Liu, Yunxin and Leng, Jingwen},
  booktitle={Proceedings of the 48th Annual ACM/IEEE International Symposium on Computer Architecture (ISCA)},
  pages={1083--1095},
  year={2021}
}

@inproceedings{jang2021sparsity,
  title={{Sparsity-aware and re-configurable NPU architecture for Samsung flagship mobile SoC}},
  author={Jang, Jun-Woo and Lee, Sehwan and Kim, Dongyoung and Park, Hyunsun and Ardestani, Ali Shafiee and Choi, Yeongjae and Kim, Channoh and Kim, Yoojin and Yu, Hyeongseok and Abdel-Aziz, Hamzah and Park, Jun-Seok and Lee, Heonsoo and Lee, Dongwoo and Kim, Myeong Woo and Jung, Hanwoong and Nam, Heewoo and Lim, Dongguen and Lee, Seungwon and Song, Joon-Ho and Kwon, Suknam and Hassoun, Joseph and Lim, SukHwan and Choi, Changkyu},
  booktitle={Proceedings of the 48th Annual ACM/IEEE International Symposium on Computer Architecture (ISCA)},
  pages={15--28},
  year={2021}
}

@inproceedings{wang2021pyramid,
  title={{Pyramid vision Transformer: A versatile backbone for dense prediction without convolutions}},
  author={Wang, Wenhai and Xie, Enze and Li, Xiang and Fan, Deng-Ping and Song, Kaitao and Liang, Ding and Lu, Tong and Luo, Ping and Shao, Ling},
  booktitle={Proceedings of the IEEE/CVF international conference on computer vision},
  pages={568--578},
  year={2021}
}

@article{sakaguchi2021winogrande,
  title={{Winogrande: An adversarial Winograd schema challenge at scale}},
  author={Sakaguchi, Keisuke and Bras, Ronan Le and Bhagavatula, Chandra and Choi, Yejin},
  journal={Communications of the ACM},
  volume={64},
  number={9},
  pages={99--106},
  year={2021},
  publisher={ACM New York, NY, USA}
}

@inproceedings{fan2023taskfusion,
  title={Taskfusion: An efficient transfer learning architecture with dual delta sparsity for multi-task natural language processing},
  author={Fan, Zichen and Zhang, Qirui and Abillama, Pierre and Shoouri, Sara and Lee, Changwoo and Blaauw, David and Kim, Hun-Seok and Sylvester, Dennis},
  booktitle={Proceedings of the 50th Annual International Symposium on Computer Architecture},
  pages={1--14},
  year={2023}
}

@inproceedings{liu2024hsconn,
  title={HSCONN: Hardware-Software Co-Optimization of Self-Attention Neural Networks for Large Language Models},
  author={Liu, Siqin and Kuve, Prakash Chand and Karanth, Avinash},
  booktitle={Proceedings of the Great Lakes Symposium on VLSI 2024},
  pages={736--741},
  year={2024}
}

@inproceedings{katharopoulos2020transformers,
  title={{Transformers are RNNS: Fast autoregressive Transformers with linear attention}},
  author={Katharopoulos, Angelos and Vyas, Apoorv and Pappas, Nikolaos and Fleuret, Fran{\c{c}}ois},
  booktitle={International conference on machine learning},
  pages={5156--5165},
  year={2020},
  organization={PMLR}
}

@inproceedings{bai2024swat,
  title={{SWAT: Scalable and efficient window attention-based Transformers acceleration on FPGAs}},
  author={Bai, Zhenyu and Dangi, Pranav and Li, Huize and Mitra, Tulika},
  booktitle={Proceedings of the 61st ACM/IEEE Design Automation Conference},
  pages={1--6},
  year={2024}
}

@article{hu2024wafer,
  title={{Wafer-scale computing: advancements, challenges, and future perspectives [Feature]}},
  author={Hu, Yang and Lin, Xinhan and Wang, Huizheng and He, Zhen and Yu, Xingmao and Zhang, Jiahao and Yang, Qize and Xu, Zheng and Guan, Sihan and Fang, Jiahao and others},
  journal={IEEE Circuits and Systems Magazine},
  volume={24},
  number={1},
  pages={52--81},
  year={2024},
  publisher={IEEE}
}

@article{wang2024tmac,
  title={{TMAC: Training-targeted mapping and architecture co-exploration for wafer-scale chips}},
  author={Wang, Huizheng and Yang, Qize and Wei, Taiquan and Yu, Xingmao and Li, Chengran and Fang, Jiahao and Lu, Guangyang and Dai, Xu and Liu, Liang and Jiang, Shenfei and others},
  journal={Integrated Circuits and Systems},
  year={2024},
  publisher={SJTU}
}

@inproceedings{wang2026WATOS,
  title     = {{WATOS: Efficient LLM training strategies and architecture co-exploration for wafer-scale chip}},
  author    = {Wang, Huizheng and Wang, Zichuan and Wang, Hongbin and Hou, Jingxiang and Wei, Taiquan and Li, Chao and Hu, Yang and Yin, Shouyi},
  booktitle = {Proceedings of the IEEE International Symposium on High-Performance Computer Architecture (HPCA)},
  year      = {2026},
  note      = {Accepted}
}

@inproceedings{wang2026TEMP,
  title     = {{TEMP: A memory efficient physical-aware tensor partition-mapping framework on wafer-scale chips}},
  author    = {Wang, Huizheng and Wei, Taiquan and Wang, Zichuan and Jiang, Dingcheng and Yang, Qize and Liu, Jiaxin and Hou, Jingxiang and Li, Chao and Deng, Jinyi and Hu, Yang and Yin, Shouyi},
  booktitle = {Proceedings of the IEEE International Symposium on High-Performance Computer Architecture (HPCA)},
  year      = {2026},
  note      = {Accepted}
}

@article{he2025waferllm,
  title={{WaferLLM: Large language model inference at wafer scale}},
  author={He, Congjie and Huang, Yeqi and Mu, Pei and Miao, Ziming and Xue, Jilong and Ma, Lingxiao and Yang, Fan and Mai, Luo},
  journal={arXiv preprint arXiv:2502.04563},
  year={2025}
}

@article{tang2025moentwine,
  title={{MoEntwine: Unleashing the potential of wafer-scale chips for large-scale expert parallel inference}},
  author={Tang, Xinru and Hou, Jingxiang and Jiang, Dingcheng and Wei, Taiquan and Liu, Jiaxin and Deng, Jinyi and Wang, Huizheng and Yang, Qize and Shang, Haoran and Li, Chao and others},
  journal={arXiv preprint arXiv:2510.25258},
  year={2025}
}

\end{document}